\documentclass[]{aa}  

\usepackage{graphicx}
\usepackage{txfonts}
\usepackage{lscape}
\usepackage[colorlinks = true,
            linkcolor = blue,
            urlcolor  = black,
            citecolor = cyan]{hyperref}

\newcommand {\kms} {\,{\rm km\,s}^{-1}}

\newcommand {\Mpc} {\,{\rm Mpc}}

\newcommand {\msun}{\,{\rm M}_\odot}

\newcommand{\Myr}{\,{\rm Myr}}
\newcommand{\Gyr}{\,{\rm Gyr}}

\newcommand{\bagpipes}{\textsc{Bagpipes}}
\def\apjl{ApJ}%
\def\apjs{ApJS}%
\def\ao{Appl.~Opt.}%
\def\aap{A\&A}%
\defcitealias{Marasco+23b}{M23}
\defcitealias{Vulcani+11}{V11}

\begin{document} 
\title{Modelling the photometric and morphological evolution of disc galaxies in the cluster environment}
\titlerunning{The evolution of disc galaxies in clusters}
\authorrunning{A. Marasco et al.}

\author{
A. Marasco\inst{1},
B.\,M. Poggianti\inst{1},
B. Vulcani\inst{1},
A. Moretti\inst{1},
M. Gullieuszik\inst{1},
J. Fritz\inst{2}
}
\institute{INAF – Padova Astronomical Observatory, Vicolo dell’Osservatorio 5, I-35122 Padova, Italy\\
\email{antonino.marasco@inaf.it}
\and
Instituto de Radioastronomia y Astrofisica, UNAM, Campus Morelia, AP 3-72, CP 58089, Mexico}
\date{Received / Accepted }

\abstract
{
Compelling observational evidence indicates that the disc population in galaxy clusters has undergone rapid evolution, transitioning from a dominance of blue spirals to red S0s over the past $\sim7\Gyr$.
We build a simplified cluster evolutionary model in the $\Lambda$CDM framework to constrain the characteristic timescales of this transformation. 
In our model, all field spirals joining the cluster at various redshifts are subject to ram-pressure stripping (RPS), which removes their gas reservoir leading to the quenching of their star formation on a timescale $t_{\rm s}$, and to an (initially) unspecified mechanism that transforms them into S0s on a timescale $t_{\rm m}$. 
We assume that $t_{\rm s}$ and $t_{\rm m}$ are independent and both power-law functions of $M_\star/M_{\rm cl}$, the galaxy-to-cluster mass ratio.
We constrain our model using the observed distribution of spirals and S0s in a color-mass plane from the OmegaWINGS and EDisCS cluster surveys at $z\simeq0.055$ and $z\simeq0.7$, respectively.
Our best-fit model reproduces the data remarkably well, especially at low redshift, and predicts evolutionary trends for the main morphological fractions in agreement with previous studies. 
We find typical $t_{\rm s}$ between $0.1$ and $1\Gyr$, compatible with previous estimates, supporting RPS as the most efficient quenching mechanisms in clusters.
A surprisingly strong anti-correlation between $t_{\rm s}$ and $M_\star/M_{\rm cl}$ is required in order to suppress the formation of red, low-mass spirals at low redshift, which we interpret as driven by orbit anisotropy.
Conversely, $t_{\rm m}$ depends very weakly on $M_\star/M_{\rm cl}$ and has typical values of a few Gyr.
The inferred morphological evolution is compatible with that resulting from the ageing of the stellar populations in galaxies abruptly quenched by ram pressure stripping: we confirm spectrophotometric ageing as a key channel for the spiral-to-S0 transition in galaxy clusters, with secular evolution playing a secondary role.
}

\maketitle
%

\section{Introduction}\label{s:introduction}

Galaxy clusters have long been recognized as key environments in which to study the impact of large-scale structure on galaxy evolution. 
As the most massive gravitationally bound systems in the Universe, clusters host deep potential wells filled with hot intracluster medium (ICM), high galaxy densities, and a complex accretion history of groups and field galaxies. 
These conditions make them natural laboratories to investigate the ensamble of external (environmental) processes that drive galaxy evolution: observations over the past decades have revealed that the galaxy population in clusters differs markedly from that in the field both in terms of morphology and star formation activity, and that these differences evolve significantly over cosmic time.

One of the most prominent signatures of cluster-driven evolution is the transformation of late-type spirals into lenticular (S0) galaxies.
Early studies of the morphology–density relation \citep{Oemler74, Dressler+80} established that local clusters are dominated by early-type systems, with spirals contributing only a minor fraction of the population. 
In contrast, later studies based on \emph{Hubble Space Telescope} (HST) observations \citep[e.g.][]{Dressler+97,Fasano+00, Postman+05,Desai+07} have shown that spirals are proportionally much more common (a factor of $\sim2-3$), and S0s much rarer, in distant (up to $z\simeq0.8$) than in nearby clusters, strongly supporting an evolutionary scenario where the cluster environment acts on newly accreted late-type galaxies, driving a morphological transition towards earlier galaxy types on timescales of a few Gyr \citep[e.g.][]{Poggianti+09, Donofrio+15}.
The main mechanism responsible for this transformation is yet to be identified, and is one of the subjects of this study.

The other key effect associated with galaxy clusters is environment-driven quenching, where star-forming spirals infalling from the field are deprived of their cold gas reservoirs, first reducing their star formation \citep{Vulcani+10, Paccagnella+16, Old+20} and eventually
evolve into passive systems. 
Amongst the various quenching mechanisms, ram-pressure stripping \citep[RPS;][]{GunnGott72} has emerged as one of the most effective one: as galaxies plunge through the dense ICM at high velocities, their cold gas is displaced and removed, often on timescales shorter than a Gyr \citep[e.g.][]{Boselli+06, Marasco+16, Jaffe+18}. 
The presence of a high fraction of post-starburst galaxies in cluster centres strongly supports RPS as a key evolutionary mechanism \citep[e.g.][]{Poggianti+99, Paccagnella+17, Paccagnella+19}.
The most spectacular manifestation of ongoing RPS is the population of so-called `jellyfish galaxies', which exhibit one-sided gaseous and stellar tails stretching tens of kpc \citep[e.g.][]{Smith+10b, Poggianti+17, Boselli+22, Poggianti+25}, providing unique snapshots of environmental quenching in action.
The observational picture across redshifts indicates that environmental quenching is already well underway by $z\simeq1$ and beyond, when clusters host a substantial fraction of passive galaxies but still retain a higher proportion of spirals compared to the present day \citep{Poggianti+09b, Webb+20, McNab+21, Baxter+22}.
This implies that morphological transformation and quenching act on different timescales, with the former process being slower than the latter, or simply becoming more efficient at later times \citep[e.g.][]{Poggianti+99, Vulcani+15}.

Observational and theoretical arguments indicate that RPS and morphological transformation in clusters are not independent processes.
Once a galaxy is deprived of its dynamically coldest component (the interstellar medium), resonances induced by spiral arms are expected to drive up the level of random motion in the stellar disc, which becomes unable to support the spiral pattern itself \citep[e.g.][and references therein]{Sellwood+14}. 
Ultimately, this `secular' evolution alters the stellar structure of the galaxy and pushes it towards the S0 type.
However, there is no consensus on the characteristic timescale of this mechanism, with values quoted in the literature ranging from $\sim2$ to $10\Gyr$ \citep[e.g.][]{SellwoodCarlberg84, Fujii+11}.

More recently, \citet[][hereafter \citetalias{Marasco+23b}]{Marasco+23b} have shown that the simple ageing of the stellar populations in a spiral galaxy that is abruptly quenched by the cluster environment leads to a spatial and spectral redistribution of its stellar light that can be sufficient to alter its morphological classification. 
Once the star formation is halted, the galaxy becomes progressively redder, more concentrated, and smoother as the blue light originating from its spiral arms fades away.
\citetalias{Marasco+23b} quantified the timescale of this transformation by building and classifying a collection of `artificially aged' synthetic images of $91$ galaxies taken from the GAs Stripping Phenomena in galaxies (GASP) project \citep{Poggianti+17}, an integral-field spectroscopic survey with MUSE@VLT aimed at studying gas removal processes in
galaxies.
They found that the morphology transformation into an S0 is completed after $1.5-3.5\Gyr$, proceeding faster in more efficient quenching scenarios.
These results have revealed a promising channel for the spiral-to-S0 transformation of cluster galaxies, providing a natural link between RPS, star formation quenching and morphological evolution.
However, to which extent this process alone is responsible for the transformation of the cluster galaxy population over the last several Gyr is yet to be determined.

The goal of this study, which complements the work of \citetalias{Marasco+23b}, is to infer the characteristic timescales of cluster-driven quenching and morphological transformation required to match the photometric and morphological properties of cluster galaxies in the redshift range $0\!<\!z\!\lesssim\!1$.
We pursue our goal by building simplified models for the evolution of cluster galaxies in the $\Lambda$CDM framework, where the main environmental mechanisms leading to galaxy quenching and morphological transformation are included in a simple, parametric form.
We compare this model with the data to infer the evolutionary timescales, which ultimately allow us to address the question of whether the spectrophotometric evolution is the main channel for the spiral-to-S0 transformation in clusters.

This paper is structured as follows.
In Section \ref{s:method} we describe our approach, providing details on the dataset used, on the evolutionary model of cluster galaxies, and on the technique adopted to compare the two.
Our results are shown in Section \ref{s:results}.
In Section \ref{s:discussion} we discuss possible variations in our modelling scheme and provide an interpretation of our results in the context of the study of \citetalias{Marasco+23b}.
A summary of this study is given in Section \ref{s:conclusion}.
Throughout this paper we adopt a flat $\Lambda$CDM cosmology with $\Omega_{\rm m,0}\!=\!0.3$ and $H_0\!=\!70\kms\Mpc^{-1}$, and a \citet{Kroupa01} initial mass function (IMF). 
Magnitudes and colours are expressed in mag units.

\section{Method}\label{s:method}
\begin{figure}
\begin{center}
\includegraphics[width=0.45\textwidth]{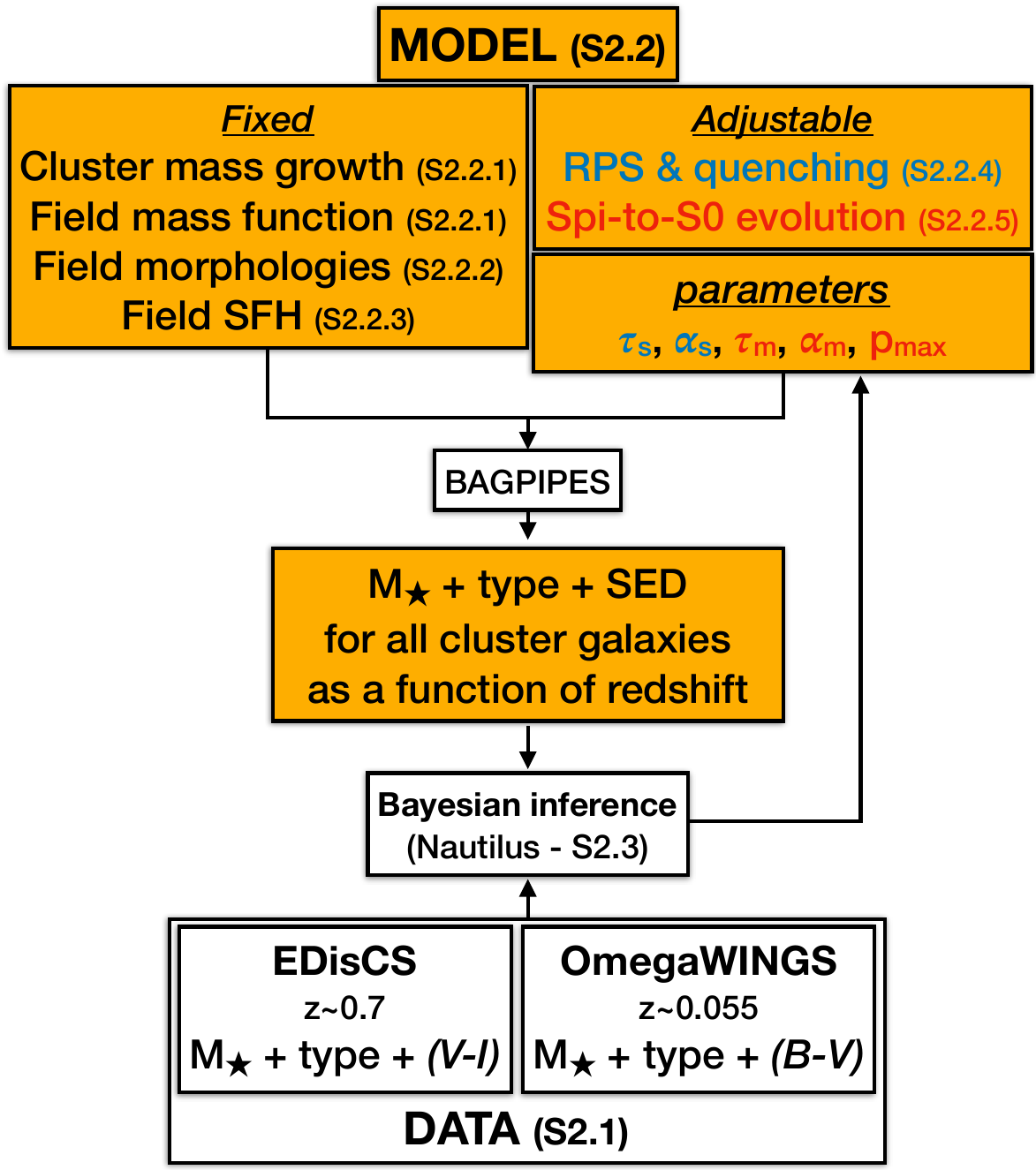}
\caption{Flowchart illustrating our method. White blocks show pre-existing dataset or software packages. Yellow blocks show the new products built for this study. The chart indicates also the Sections of the main text that provide details on a given part of the method.}
\label{f:method}
\end{center}
\end{figure}

Our method is based on the modelling of the photometric and morphological properties of the cluster galaxy population at different redshifts.
The model, detailed in Section \ref{ss:model}, follows the build-up of cluster galaxies as a function of redshift and is engineered to predict the $B$, $V$ and $I$-band magnitudes and the main morphology types (spiral--Spi, S0, elliptical--Ell) of cluster galaxies. 
The parameters of the model are constrained using observational data, detailed in Section \ref{ss:data}.
In Fig.\,\ref{f:method} we provide a flowchart to guide the reader through our method.

\subsection{Data}\label{ss:data}
The two cluster samples used in this study, OmegaWINGS \citep{Fasano+06,Gullieuszik+15,Moretti+14,Moretti+17} and EDisCS \citep{White+05}, have three key properties that make them the optimal choice for our investigation.
First, they sample two redshift bins that feature a substantial difference in the morphology distribution of the cluster galaxy population \citep[e.g.][]{Fasano+00,Vulcani+11, Vulcani+11b, Vulcani+14}, supporting the need for a significant cluster-driven evolution over the last several Gyr. 
Second, as shown in detail in Section \ref{sss:cluster_growth}, the typical cluster properties are such that the EDisCS clusters can be considered as progenitor of the OmegaWINGS clusters \citep{Vulcani+14}.
Third, the two samples have compatible morphological classification scheme, as discussed below and in \citet[][hereafter \citetalias{Vulcani+11}]{Vulcani+11}.
These properties ensure that the OmegaWINGS and EDisCS data provide homogeneous and highly significant information to constrain our evolutionary models. 

OmegaWINGS is a survey of 46 nearby ($0.04\!<\!z\!<\!0.07$) galaxy clusters and of their peripheries, imaged in the \emph{B} and \emph{V} bands with OmegaCAM \citep{Kuijken11} at the VLT Survey Telescope \citep[VLS;][]{CapaccioliSchipani11}.
Each cluster is covered beyond its virial radius $R_{200}$, extending the environment range covered by the parent WINGS Survey \citep{Fasano+06} and permitting to image the surrounding infall region.
Spectroscopic follow-ups for 36 OmegaWINGS clusters were taken with the AAOmega spectrograph \citep{Smith+04,Sharp+06} at the Australian Astronomical Observatory (AAT) for redshift measurements and assignment of cluster memberships \citep{Moretti+17}.
The morphological classification of OmegaWINGS galaxies has been performed using the MORPHOT package \citep{Fasano+12} by \citet{Vulcani+23}, and is based on 21 morphological diagnostics, computed directly from the \emph{V}-band images, to provide two independent classifications (one based on a maximum likelihood technique, and the other one on a neural network machine) which in turn are used to infer the morphological type $T_{\rm M}$.
Here we focus on spectroscopically confirmed galaxies within $1\times R_{200}$ cluster-centric distance and with robust $T_{\rm M}$ measurements, and assign a simple morphology label to each galaxy (Ell: $-5.5\!<\!T_{\rm M}\!<\!-4.25$, S0: $-4.25\!\leq\!T_{\rm M}\!\leq\!0$, Spi\footnote{Irregulars are included in this category.}: $T_{\rm M} \!>\!0$).
Our final OmegaWINGS sample is made of $7162$ galaxies ($2358$ Spi, $3201$ S0, $1603$ Ell), complete down to $m_V\!=\!20$ and with a median redshift of $0.055$, which we take as the characteristic redshift of this sample.
We stress that the morphological mix remains steady across the redshift range spanned by the sample, in spite of the spatial resolution varying by a factor of $\sim1.7$ from $z\!=\!0.04$ to $z\!=\!0.07$. 
$M_\star$ estimates are taken from \citet{Vulcani+22}, and are based on the relation between $M_\star/L_{\rm B}$ and the rest-frame ($B-V$) colour from \citet{BelldeJong01}, converted to a \citet{Kroupa01} IMF.

EDisCS is a multiwavelength photometric and spectroscopic survey of galaxies in clusters at $0.4\!<\!z\!<\!1$, based on 20 fields selected from Las Campanas Distant Cluster Survey (LCDCS) catalogue \citep{Gonzalez+01}. 
For all 20 fields deep optical multiband photometry obtained with FORS2/VLT \citep{White+05} and near-IR photometry obtained with SOFI/NTT is available. 
FORS2/VLT was additionally used to obtain spectroscopy for $19$ of the fields \citep{Halliday+04,Milvang-Jensen+08,Vulcani+12}, covering up to $R_{200}$ for all clusters except one.
For consistency with WINGS/OmegaWINGS, $M_\star$ estimates were re-derived by \citetalias{Vulcani+11} using the \citet{BelldeJong01} approach, with total magnitudes inferred from photo-$z$ fits \citep{Pello+09} and rest-frame luminosities determined with the method of \citet{Rudnick+09}.
The EDisCS morphological classification is based on visual inspection of the ACS/HST mosaic F814W imaging, acquired for $10$ of the highest redshift clusters \citep{Desai+07}.
The resulting sample is made by $254$ galaxies ($156$ Spi, $45$ S0, $53$ Ell) with $m_I\lesssim22$ and a median redshift of $\sim0.7$, which we use as a characteristic redshift for this sample. 
We found no significant differences in the morphological mix of EDisCS galaxies above and below $z=0.7$, in spite of the F814W filter sampling different rest-frame wavebands across the sample.

Our investigation requires that the high- and low-$z$ galaxy samples have compatible morphology estimates.
This was tested in \citetalias{Vulcani+11}, where three of the authors originally performing the EDisCS classification provided a visual classification for a subsample of the WINGS galaxies. 
They found a very good agreement between the visual and the MORPHOT morphological types, with discrepancies compatible with those emerging from comparing the three visual classifications amongst themselves.
We notice that, at the typical $z$ of the EDisCS galaxies, the F814W band samples the the rest-frame V-band and provides a spatial resolution (in kpc) comparable with the OmegaCAM images of OmegaWINGS clusters, which is key for a consistent classification between the two samples.

For this study, we use the EDisCS and OmegaWINGS datasets to derive the distribution of Spi and S0 galaxies in a color-$M_\star$ plane, using the observed-frame ($B-V$) colour for OmegaWINGS and the ($V-I$) colour for EDisCS.
We focus on a mass range for which OmegaWINGS is expected to be complete ($M_\star\gtrsim10^{9.8}$) but we anticipate that, in our comparison with the models, we will normalise the predicted $M_\star$ distributions for Spi and S0 to those observed in order to minimise issues due to mass incompleteness (see Section \ref{ss:fitting_tecnique}).
We account for spectroscopic incompleteness in OmegaWINGS and EDisCS using the available correction weights $w$ associated with each cluster member\footnote{$w\!\equiv\!1/R$ (with $0\!<\!R\!\leq\!1$), where $R$ is derived as the ratio of number of spectra yielding a redshift to the total number of galaxies in the parent photometric catalogue \citep{Poggianti+06, Paccagnella+16, Moretti+17}. We exclude galaxies with $R\!<\!0.2$ as they are excessively under-represented in the observed sample.}, although we stress that this correction does not impact our results.
The resulting 2D distributions are shown later on in Fig.\,\ref{f:hist2d_results} and discussed in Section \ref{s:results}, and are used to constrain our evolutionary model, which we describe in detail below.

\subsection{Model}\label{ss:model}
\subsubsection{Cluster growth}\label{sss:cluster_growth}
\begin{figure}
\begin{center}
\includegraphics[width=0.48\textwidth]{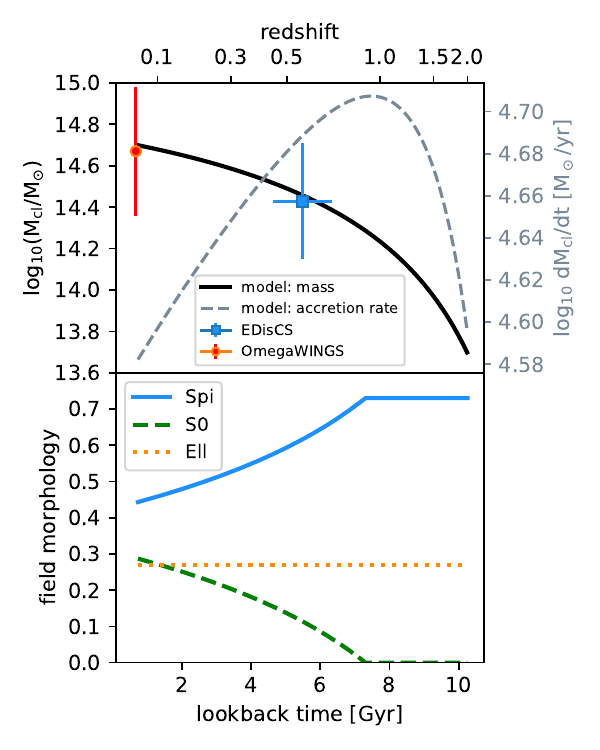}
\caption{Evolutionary properties of our model cluster. \emph{Top panel}: mass (solid black curve) and accretion rate (dashed grey curve) vs redshift for a cluster with $M_{\rm cl}\!=\!5\times10^{14}\msun$ at $z\!=\!0.055$. The blue square and the red circle show the median $M_{\rm cl}$ and redshifts of the EDisCS and OmegaWINGS clusters, respectively, with the error bars corresponding to half the difference between the 84th and 16th percentiles of the galaxy distribution. \emph{Bottom panel}: morphology fractions for field galaxies that join the cluster as a function of $z$, assumed to be independent of the galaxy $M_\star$.}
\label{f:halo_growth}
\end{center}
\end{figure}
The first step in our modelling procedure is to determine cluster mass growth rates as a function of time.
This is achieved using the halo growth model of \citet{Correa+15a}\footnote{available at \url{https://www.camilacorrea.com/code/commah/}}, based on the extended \citet{PressSchechter74} formalism. 
This model provides a very good match to the halo accretion histories inferred from more detailed semi-analytical models of galaxy evolution, such as the Millennium \citep{McBride+09} and the Bolshoi \citep{vandenBosch+14} suites.
The model is fully constrained by assuming a specific halo mass at given redshift: we focus on a single cluster with a mass $M_{\rm cl}$ of $5\times10^{14}\msun$ at $z\!=\!0.055$, which are the median mass and redshift of OmegaWINGS clusters\footnote{In this work we assume $M_{\rm cl}\equiv M_{\rm 200}$, the mass within a radius $R_{200}$ within which the mean dark matter density is $200$ times the critical density of the Universe.}. 

We compute the growth history of this halo from $z\!=\!2$ to $z=0.055$ and show the results in the top panel of Fig.\,\ref{f:halo_growth}. 
$M_{\rm cl}$ has grown by a factor of $\sim2$ from $z=0.7$ to $z=0.055$.
We infer $M_{\rm cl}$ for the EDisCS and OmegaWINGS clusters from their $R_{200}$ estimates \citep{Vulcani+13,Biviano+17}, and plot in Fig.\,\ref{f:halo_growth} the median value of their distributions, together with the difference between the 84th and 16th percentiles (blue square and red circle with error-bars).
Clearly, there is excellent agreement between our growth model and the typical EDisCS and OmegaWINGS cluster masses: this allows us to treat our high- and low-redshift data within a common evolutionary framework \citep[see also][]{Vulcani+14c}.
Additionally, Fig.\,\ref{f:halo_growth} shows that $M_{\rm cl}$ at $z\!=\!2$ was about factors $5$ and $10$ lower than at $z=0.7$ and $z=0.055$, respectively.
This is a key aspect of our method because it ensures that the modelled cluster galaxy population at $z=0.7$ and $z=0.055$ will be dominated by systems acquired from the field and evolved according to the recipes implemented, rather than by the initial (proto)cluster population at $z\simeq2$, whose properties are largely unconstrained.
In other words, at the redshifts where the model will be compared with the observations, it will have `lost memory' of its (poorly constrained) initial conditions.

As a next step, we populate the halo with galaxies.
These are injected within the cluster from the field in time-steps of $0.1\Gyr$, following the redshift-dependent galaxy stellar mass function of \citet{McLeod+21}.
We assume that the cluster galaxy mass growth rate is proportional to the halo mass growth rate: that is, if in the time interval from $t$ to $t+\Delta t$ the halo mass has increased by $\Delta M_{\rm cl}\equiv\dot{M}_{\rm cl}\Delta t$, additional $\Delta M_\star\equiv f_\star \Delta M_{\rm cl}$ will be added to the cluster, randomly sampled from the assumed galaxy mass function at redshift $z(t)$.
In principle, $f_\star$ can be chosen so that the final number of cluster galaxies above a given $M_\star$ (or below a given magnitude) is similar to that of the typical OmegaWINGS cluster.
However, as we are not interested in modelling individual clusters but focus on the properties (mass, colour and morphological mix) of the mean galaxy population, we can arbitrarily inflate $f_\star$ to ensure sufficiently large statistics. 
In particular we adopt $f_\star=0.6$, which for our best-fit model (Section \ref{ss:data_vs_model}) provides  $\sim3\times10^3$ disc galaxies (Spi+S0s) with $m_{\rm I}<22$ at $z\!=\!0.7$, and $\sim10^4$ with $m_{\rm V}<20$ at $z\!=\!0.055$.
These are factors $5\!-\!10$ larger than the number of discs in the EDisCS and OmegaWINGS datasets, ensuring a robust statistical characterisation of the model galaxy population.

\subsubsection{Field morphology}\label{sss:field_morphology}
Our model requires to assign an initial morphology to all galaxies that join the cluster from the field.
Ideally, we would need the morphology distribution of field galaxies as a function of $M_\star$ and $z$, which we would use to randomly assign an initial type to all galaxies in the model.
This is a challenging request: while observational constraints are numerous both in the local Universe \citep[e.g][]{NairAbraham2010,WilmanErwin12,Moffett+16,DS+18} and at high redshift \citep[e.g.][]{vanderWel+07, Oesch+10, Huertas-Company+16, Cavanagh+23, Kartaltepe+23},
they are somewhat inhomogeneous and frequently in tension with one another.
Our strategy is to focus on field morphology estimates that are compatible with those adopted for our cluster galaxies.
We rely on the study of \citet{Calvi+12}, who analysed a sample of $\sim2000$ systems from the Padova–Millennium Galaxy and Group Catalogue \citep[PM2GC;][]{Calvi+11}, using MORPHOT (as in OmegaWINGS) to assign morphological types.
For their mass-limited sample ($M_\star>10^{10.2}\msun$), \citet{Calvi+12} found that the morphological mix shows only a weak dependence on $M_\star$ and on the environment (with the exception of clusters), resulting in $27.0\%$ of Ell, $28.7\%$ of S0s, and $44.3\%$ of Spi in the so-called `general field' environment (PM2-GF), which collects galaxies in groups, binary systems, and in isolation. 
The general field is built to represent the typical galaxy population that accretes onto nearby clusters: as most galaxies in the PM2-GF belong to groups, we expect pre-processing effects to be accounted for.
We use the percentages quoted above, valid for all $M_\star$, to characterise the field population in our lowest redshift bin.
For simplicity we do not use the estimated uncertainties on these percentages, which in any case are small ($1.3-1.5\%$).

The extrapolation of these fractions to higher redshifts requires additional assumptions.
As our model does not treat the morphological evolution of ellipticals, the final Ell fraction will depend solely on that accreted from the field.
There is observational evidence for an approximately constant $\sim30\%$ Ell fraction in clusters at least up to $z\simeq0.6$ \citep[][\citetalias{Vulcani+11}]{Fasano+00, Vulcani+23}: to reproduce this result, we assume that the Ell fraction adopted for the low-$z$ field is valid at all $z$, which guarantees a $27\%$ of Ell in the model cluster at all epochs.
A convenient outcome of this assumption is that we can safely factor out the ellipticals from our modelling scheme, and focus solely on the evolution of the disc (Spi+S0) population. 
Conversely, several studies agree on the tendency for the field Spi fraction to increase with $z$ \citep[][]{Dressler+97,Bundy+05,Bundy+06,Conselice14}, although there is no consensus on the exact trend. 
Here we follow the results of \citet[][see their Fig.\,7]{Cavanagh+23} and set the Spi fraction to increase linearly with $z$ so that, by $z\simeq0.9$, the disc (Spi+S0) field population is fully dominated by spirals.
As this assumption has visible impacts on our results but is only marginally supported by the observational data, we discuss alternative scenarios in Section \ref{ss:field_extreme}.

The bottom panel of Fig.\,\ref{f:halo_growth} shows the assumed evolution of the field morphological fractions. 
The field fractions at $z\!=\!2$ are also used to define the initial morphological mix within the (proto)cluster.
This has virtually no impact on our results because, as discussed in Section \ref{sss:cluster_growth}, the cluster initial conditions are rapidly diluted during the evolution.

\subsubsection{Star formation histories and model photometry}\label{sss:SFH_photo}
Detailed modelling of galaxy photometry, and in particular of galaxy colours, is key to anchor our models to the data.
In order to have complete control over the photometric evolution of the model galaxies, we have chosen to fully characterise their spectral energy distribution (SED) using the Bayesian Analysis of Galaxies for Physical Inference and Parameter EStimation (\bagpipes) software package \citep{Carnall+18}, adopting a series of assumptions that we discuss below. 

The main ingredient to build the SED of a galaxy is its star formation history (SFH).
We constrain the SFH of field spirals by assuming that, until the time at which they join the cluster, they have being lying on the (redshift-evolving) main sequence (MS) of star formation.
The reference MS model that we adopt here is that of \citet{Popesso+23}, but in Appendix \ref{a:supplementary} we show that the MS of \citet{Speagle+14} leads to compatible results.
Following \citet{Gladders+13}, we parametrise the galaxy SFHs using a log-normal function, which is known to describe well the evolution of the specific SFR of galaxies with time and the overall star formation rate density of the Universe. 
In Appendix \ref{a:SFH_MS} we detail how to derive a relation between the present-day $M_\star$ of a galaxy and the two free parameters of its log-normal SFH so that galaxy remains approximately on the MS for $0<z<2$.
As shown in Appendix \ref{a:SFH_MS}, galaxy downsizing arises naturally under these assumptions.

\bagpipes\ uses the \citet{BruzualCharlot03} stellar population synthesis (SPS) models (2016 version) with the \citet{KroupaBoily02} initial mass function.
$M_\star$ estimates include stellar remnants but exclude the gas lost by stellar winds and supernovae.
We assign a unique value for the gas-phase metallicity of any galaxy based on its $M_\star$ and SFR using the fundamental metallicity relation (FMR) of \citet{Curti+20}, assumed to be redshift-independent \citep[e.g.][]{Mannucci+10}.
We adopt the dust attenuation model of \citet{CF00}, using a value of $0.7$ for the power law slope of the attenuation law and a multiplicative factor of $3$ for the $A_{\rm V}$ associated with the birth clouds (which in \bagpipes\ have a lifetime of $10\Myr$).
A visual extinction $A_{\rm V}$ is assigned to each field spiral following the prescription of \citet{GarnBest10}, which is based on the galaxy $M_\star$ alone.
\bagpipes\ is used to infer observed- and rest-frame apparent and absolute magnitudes in the $B$, $V$ and $I$ bands, using the appropriate OmegaCAM and FORS2 broad-band filters to best match the observed photometric measurements.

\subsubsection{RPS and cluster-driven quenching}\label{sss:model_quenching}
\begin{figure}
\begin{center}
\includegraphics[width=0.48\textwidth]{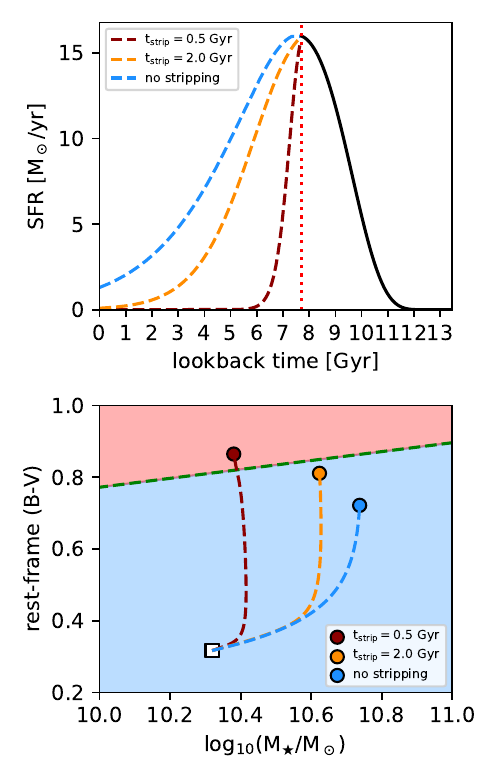}
\caption{Photometric evolution and mass build-up of an idealised main-sequence galaxy with initial $M_\star\simeq2\times10^{10}\msun$ that joins our cluster model at $z\!=\!1$, for stripping timescales of $0.5$ and $2.0\Gyr$ (dark-red and orange curves, respectively), and for a case without stripping (blue curve). \emph{Top panel}: SFR as a function of the lookback time. For times larger than $\sim8\Gyr$ ($z>1$), the three models share a common SFH shown by the solid black curve. \emph{Bottom panel}: galaxy evolution in the rest-frame ($B-V$) colour vs $M_\star$ plane. The white square shows the initial values at $z\!=\!1$, the coloured circles show the final ($z\!=\!0$) values for the three scenarios considered. The green dashed-line separates the red sequence from the blue cloud \citep[from][]{Vulcani+22}.}
\label{f:photo_track}
\end{center}
\end{figure}

One of the main photometric features of the cluster galaxy population is the presence of a well populated red sequence, built by galaxies of various morphological types whose star formation has been halted.
We model environmental quenching in a scenario where its main physical driver is RPS from the ICM, which removes the galaxy's ISM, hence inhibits the star formation processes.
As in \citetalias{Marasco+23b}, the stripping proceeds outside-in at an exponential rate, reducing the radius of the ISM disc as
\begin{equation}\label{eq:stripping}
    R(t) = R_{\rm max}\,\exp(-t/t_{\rm s}),
\end{equation}
where $t$ is the time since the galaxy has joined the cluster, $R_{\rm max}$ is the disc radius at $t=0$, and $t_{\rm s}$ is the stripping timescale.
This exponential form is meant to capture qualitatively the increasing stripping efficiency affecting a galaxy in acceleration within a progressively denser ICM, and allows for a direct comparison with the results of \citetalias{Marasco+23b}, discussed in Section \ref{ss:rps_evolution}.
We can compute analytically how the SFR drops in this stripping scenario under the assumptions of pure exponential discs and that the SFR density radial profile is proportional to the gas surface density profile \citep[which implies a unitary slope in the Kennicutt-Schmidt relation, as suggested by][]{Leroy+13}.
We find
\begin{equation}\label{eq:SFRdrop}
    \frac{\rm{SFR}(t)}{\rm{SFR}(0)} = \frac{1-e^{-\delta(t)} \left[1+\delta(t)\right]}{1-e^{-\delta(0)} \left[1+\delta(0)\right]}\,, 
\end{equation}
where $\delta(t)\equiv R(t)/R_{\rm d}$, with $R(t)$ given by Equation\,(\ref{eq:stripping}) and $R_{\rm d}$ being the exponential scale length of the disc.
We can remove the dependence on $R_{\rm d}$ by imposing an initial disc radius proportional to the scale length: specifically, we set $R_{\rm max}\!=\!5\times R_{\rm d}$, a size that encloses $\sim95\%$ of the mass for an exponential profile. 
Consequently, the drop in SFR modelled by Equation\,(\ref{eq:SFRdrop}) is regulated by $(t/t_{\rm s})$ alone. 
As a reference, from the moment the galaxy joins the cluster, the SFR drops to $\sim60\%$ of its initial value after $t_{\rm s}$, to $\sim15\%$ after $2\times t_{\rm s}$, and to $\sim3\%$ after $3\times t_{\rm s}$.

The stripping timescale is one of the main parameter of our model.
We assume a scenario a-la \citet{GunnGott72}, where the stripping efficiency is regulated by the competition between the gravitational restoring force of the galaxy, which (at least in the proximity of the disc) depends on $M_{\star}$, and the cluster gas density and velocity dispersion, both of which depend on $M_{\rm cl}$.
In this scenario, we can parametrise $t_{\rm s}$ as a power-law function of the galaxy-to-cluster mass ratio:
\begin{equation}\label{eq:ts}
    \log_{10}\left(\frac{t_{\rm s}}{\rm Gyr}\right) = \log_{10}\left(\frac{\tau_{\rm s}}{\rm Gyr}\right) + \alpha_{\rm s}\left[\log_{10}\left(\frac{M_\star}{M_{\rm cl}}\right)+4.7\right]
\end{equation}
where $\tau_{\rm s}$ and $\alpha_{\rm s}$ are free parameters.
The $4.7$ addend on the right-hand side of Equation\,(\ref{eq:ts}) is chosen so that $\tau_{\rm s}$ is the stripping timescale for a galaxy with $\log_{10}(M_\star/\msun)\!=\!10$ in a cluster with $\log_{10}(M_{\rm cl}/\msun)\!=\!14.7$, the typical $M_{\rm cl}$ of the OmegaWINGS clusters.

We stress that Equation\,(\ref{eq:ts}) does not come from a rigorous treatment of the RPS physics in galaxy clusters, but it is a rather simplified and convenient way to outline the trends between $t_{\rm s}$, $M_\star$ and $M_{\rm cl}$ based on our physical intuition.
We notice that the RPS efficiency is expected to depend also on the cluster concentration, $c_{\rm cl}$. 
However, in the framework assumed, $c_{\rm cl}$ can be expressed as a power-law function of $M_{\rm cl}$ using the halo concentration-mass relations \citep[e.g.][]{Dutton&Maccio14}, thus such dependency is effectively reabsorbed in the $\alpha_{\rm s}$ parameter of Equation\,(\ref{eq:ts}).

We track the SFH of galaxies subject to this cluster-driven quenching and update their model photometry using the same approach of Section \ref{sss:SFH_photo}.
Similarly to \citetalias{Marasco+23b}, we include the effect of shock-driven dust destruction by lowering the extinction $A_{\rm V}$ by a factor $\exp[-t/(t_{\rm s}+t_{\rm dust})]$, with $t_{\rm dust}=100\,\Myr$ \citep{Jones+94, Jones+96}.
Dust destruction makes galaxies bluer, marginally slowing down their pathway to the red sequence.
Also, we consider the stellar metallicity frozen to the value the galaxy had when it joined the cluster.
This assumption simplifies our calculations and is justified by the mild ($\sim0.1$ dex) increase in (light-weighted) stellar metallicity that quiescent galaxies show with respect to active galaxies of similar $M_\star$ and redshift \citep[e.g.][]{Gallazzi+25}.

In Fig.\,\ref{f:photo_track} we illustrate the case of a representative MS galaxy with initial $M_\star\simeq2\times10^{10}\msun$ that joins the cluster at $z\!=\!1$.
We follow the mass build-up and the photometric evolution of this system down to $z\!=\!0$ for three different scenarios: fast stripping with $t_{\rm s}=0.5\Gyr$, slow stripping with $t_{\rm s}\!=\!2.0\Gyr$, and no stripping at all. 
The top panel of Fig.\,\ref{f:photo_track} shows that, even in the no-stripping scenario, the galaxy SFR has dropped by a factor $\sim8$ since $z\!=\!1$, in agreement with the evolution of the MS and of the cosmic SFR density \citep[e.g.][]{MadauDickinson14, Enia+22}.
Regardless of the scenario considered, the galaxy rest-frame ($B-V$) increases with time (bottom panel of Fig.\,\ref{f:photo_track}) due to the combined effect of the ageing of the stellar populations and of the overall decline of the SFR.
The no-stripping scenario is the one that provides the bluest and most massive system at $z\!=\!0$, as expected from galaxies that evolve on the main sequence of star formation.
The fast stripping model produces the reddest and least massive system - in fact, it is the only scenario that brings the galaxy on the red sequence - whereas the slow stripping leads to masses and colours that are in between the other two.
These illustrative experiments already indicate that short stripping timescales are needed to build a well populated red-sequence of cluster galaxies.

S0s that join the cluster from the field are treated differently, under the assumption that they are already quenched galaxies and will evolve passively within the cluster environment.
We assume that they were MS galaxies in the past, and that their quenching had begun exactly $\Delta t_{\rm S0}\!=\!1\Gyr$ before joining the cluster, occurring on very short time scales equivalently to $t_{\rm s}=0.1\Gyr$ in our model.
Although observed early type galaxies show a large variety in their quenching epochs and timescales \citep[e.g.][]{Tacchella+22}, the idealised quenching scenario adopted here is meant to produce galaxies that, by the time they enter the cluster, are already `red-and-dead' regardless of their specific quenching channel. 
Increasing $\Delta t_{\rm S0}$ and/or further reducing $t_{\rm s}$ have negligible impacts on the colours of our field S0 population, which is the main observable used in our analysis.

\subsubsection{Cluster-driven morphological transformation}\label{sss:model_morphevo}
We finally introduce the cluster-driven morphological transformation in our models.
Conversely to what we have assumed for the quenching, the approach that we adopt here is agnostic of the exact mechanism responsible for the transformation: our goal is to determine whether a transformative mechanism is needed to begin with and, if so, what is its characteristic timescale.

We treat the transformation with a statistical approach, where spirals that enter the cluster will progressively increase their probability $p_{\rm S0}$ of evolving into an S0 up to a maximum probability value $p_{\rm max}$.
We parametrise $p_{\rm S0}$ as: 
\begin{equation}\label{eq:pS0}
p_{\rm S0}(t) = p_{\rm max}[1-\exp(-t/t_{\rm m})] 
\end{equation}
where $t_{\rm m}$ is the morphological transformation timescale. 
We notice that the evolution is fully suppressed for $p_{\rm max}\!=\!0$.
We impose $p_{\rm S0}\!=\!1$ for all S0 field galaxies, for which we assume no morphological evolution within the cluster environment.

Although the parametric form chosen for $p_{\rm S0}$ mimics the trends found by \citetalias{Marasco+23b} for their spectrophotometric evolution of quenched galaxies (see, for instance, their Fig.\,5), it is general enough to be applicable to any evolutionary channel. 
We assume a unique value of $p_{\rm max}$ for all galaxies at all redshifts, and consider it as a free parameter of the model.
By analogy with our treatment of the stripping timescale (Section \ref{sss:SFH_photo}), we assume that $t_{\rm m}$ is a power-law function of $M_\star/M_{\rm cl}$:
\begin{equation}\label{eq:tm}
    \log_{10}\left(\frac{t_{\rm m}}{\rm Gyr}\right) = \log_{10}\left(\frac{\tau_{\rm m}}{\rm Gyr}\right) + \alpha_{\rm m}\left[\log_{10}\left(\frac{M_\star}{M_{\rm cl}}\right)+4.7\right]
\end{equation}
where $\tau_{\rm m}$ and $\alpha_{\rm m}$ are additional free parameters of the model.
In practice, in our model $p_{\rm max}$ can be considered as a regulator of the global efficiency of the morphological transformation, irrespectively of the galaxy or the cluster masses, which instead regulate the transformation timescales.

We notice that, in our agnostic approach, we consider $t_{\rm s}$ and $t_{\rm m}$ as separate, independent timescales. 
However, in a scenario where the morphological transition follows from the galaxy quenching, we expect that $t_{\rm m}\!>\!t_{\rm s}$.
We prefer not to force this inequality in our models, and expect it to arise naturally as a data-driven result. 

\subsection{Fitting technique}\label{ss:fitting_tecnique}

\begin{table}
\caption{Model parameters.}
\label{t:param_Popesso}
\centering
\begin{tabular}{lcccc}
\hline\hline
\noalign{\smallskip}
param. & ref. & prior & fiducial & best-fit \\
\hline
\noalign{\smallskip}
\multicolumn{5}{c}{cluster-driven quenching (RPS)}\\
\hline
\noalign{\smallskip}
$\log_{10}(\tau_{\rm s}/\rm Gyr)$ & Eq.\,(\ref{eq:ts}) & [-1.3,1.3] & $-0.35^{+0.18}_{-0.28}$ & $-0.15$\\
\noalign{\smallskip}
$\alpha_{\rm s}$ & Eq.\,(\ref{eq:ts}) & [-2.0,2.0] & $-0.88^{+0.45}_{-0.61}$ & $-0.65$\\
\hline
\noalign{\smallskip}
\multicolumn{5}{c}{cluster-driven morphological evolution}\\
\hline
\noalign{\smallskip}
$\log_{10}(\tau_{\rm m}/\rm Gyr)$ & Eq.\,(\ref{eq:tm}) & [-1.3,1.3] & $0.48^{+0.20}_{-0.33}$ & $0.76$\\
\noalign{\smallskip}
$\alpha_{\rm m}$ & Eq.\,(\ref{eq:tm}) & [-2.0,2.0] & $-0.22^{+0.35}_{-0.46}$ & $-0.20$\\
\noalign{\smallskip}
$p_{\rm max}$ & Eq.\,(\ref{eq:pS0}) & [0.0,1.0] & $0.80^{+0.13}_{-0.12}$ & $1.00$\\
\noalign{\smallskip}
\hline
\multicolumn{5}{p{0.45\textwidth}}{\small {\bf Column description:} (1)--parameter, (2)--reference Equation in the text, (3)--prior range, assuming uniform distributions, (4)--fiducial values from the 16th, 50th and 84th percentiles of the marginalised posterior distribution, (5)--values that maximise the likelihood.}
\end{tabular}
\end{table}

Our model uses five free parameters, listed in Table \ref{t:param_Popesso}.
Two regulate the cluster-driven stripping timescales ($\tau_{\rm s}$, $\alpha_{\rm s}$), and three regulate the cluster-driven morphological evolution ($\tau_{\rm s}$, $\alpha_{\rm s}$, $p_{\rm max}$).
These are constrained using the Spi and S0 distribution in two separate colour-$M_\star$ planes (see Section \ref{ss:data} and Fig.\,\ref{f:hist2d_results}), which our model can predict at any $z$ for any combination of its free parameters, using the $p_{\rm S0}$ (Equation \ref{eq:pS0}) associated with each galaxy as a weight to infer the S0 fraction.

These distributions are built in the same way for the model and the data: we consider the mass range $9.75\!<\!\log_{10}(M_\star/M_\odot)\!<\!11.25$ in bins of $0.25$ dex, the (observed-frame) colour range $0.3\!<\!(B-V)\!<\!1.5$ in bins of $0.1$ mag for the model at $z\!=\!0.055$ and for the OmegaWINGS data, and the colour range $0.8\!<\!(V-I)\!<\!3.0$ in bins of $0.2$ mag for the model at $z\!=\!0.7$ and for the EDisCS data.
The use of larger colour bins for EDisCS is justified by the lower number of galaxies available compared with OmegaWINGS.
For simplicity, in our models we focus solely on the two time `snapshots' that best match the median redshifts of the two surveys, but we have verified that the use of broader redshift distributions that approximately match those observed does not alter our results.

Before building the 2D distributions, some adjustments to the model are needed in order to make the comparison with the data meaningful.
First, we consider only galaxies in our model with $m_V<20$ at $z=0.055$, and with $m_I<22$ at $z=0.7$, to account for the magnitude completeness of the observed samples.
Second, we weight our model galaxies so that the disc (Spi+S0) galaxy $M_\star$ function in the model matches the observed ones.
This means that, by construction, the total number of model discs galaxies in any bin of $M_\star$ will match the data perfectly, but the colour and morphological mix will still depend on the model parameters.
This step allows us to circumvent potential issues associated with mass incompleteness in the data and/or with an incorrect mass function in the model, and is required to implement the metric discussed below.
Finally, we must account for measurement errors that broaden the observed galaxy distribution in the color-$M_\star$ plane.
This is done by introducing random (Gaussian) fluctuations in the colour and $M_\star$ of the model galaxy population.
We assume a $0.3$ dex scatter in $M_\star$ based on a comparison between the \citet{BelldeJong01} method and $M_\star$ estimates from full SED modelling \citepalias[e.g.][]{Vulcani+11}.
Following \citet{Marasco+25L}, we assume a minimum uncertainty of $5\%$ on all flux measurements, which is added (in quadrature) to the nominal photometric uncertainty leading to a final scatter between $0.05$ and $0.07$ mag (depending on the band) on our model photometry.
 
The metric adopted to quantitatively evaluate the differences between the data and the model, both of which are in the form of galaxy number counts in bins of colour and $M_\star$, is based on Poisson statistics. 
Following \citet{Bonamente+25}, we write the Poisson likelihood as:
\begin{equation}\label{eq:loglk}
    \ln \mathcal{L}(\mathbf{x}) = k\sum_{i=1}^N \left[ D_i \, \ln M_i(\mathbf{x}) - M_i(\mathbf{x}) - \ln \left( D_i! \right) \right]\ ,
\end{equation}
where $\mathbf{x}$ are the free parameters of the model, $D_i$ and $M_i$ are data and model galaxy number count in the $i$-th bin of the colour-$M_\star$ plane, the sum is extended to all $N$ bins considered, and $0\!<\!k\!\le\!1$ (see below).
We compute Equation\,(\ref{eq:loglk}) four times - for the two redshift bins and for the Spi and S0 populations separately - and sum the resulting likelihoods together to obtain the total one.

In an idealised scenario where $\mathbf{x}$ can be adjusted so that the difference between the model and the data are driven by Poissonian errors alone, one should set $k=1$ in Equation\,(\ref{eq:loglk}). 
However, considering all the assumptions and simplifications done in our approach, this scenario is not realistic and the typical differences between data and model will be larger. 
Lowering $k$ allows to compensate for this, artificially boosting the uncertainties in the data and avoiding unrealistically small error-bars in the model parameters.
We set $k$ to an ad-hoc value of $0.03$, chosen so that models whose parameters differ by $\pm1\sigma$ rms from the best-fit values lead to visually appreciable differences in the 2D distributions.

We fit the model to the data in the context of Bayesian statistics using the Nautilus sampler \citep{Lange+23} to infer the posteriors, using uniform priors in the ranges reported in Table \ref{t:param_Popesso} and the likelihood discussed above.
We distinguish between the `best-fit' parameters, which are those that maximise our likelihood, and the `fiducial' parameters, given by the median of the marginalised posterior distributions. 
Upper and lower uncertainties on the fiducial parameters are provided by the 84th and 16th percentiles of the marginalised posteriors.

\section{Results}\label{s:results}
We now show the comparison between the data and our best-fit model (Section \ref{ss:data_vs_model}), present our predictions for the RPS and morphological transformation timescales (Section \ref{ss:timescales}), and for the evolution of the morphological mix in clusters in the latest $\sim6\Gyr$ (Section \ref{ss:evolution_morphmix}).
\subsection{Data vs best-fit model} \label{ss:data_vs_model}
\begin{figure*}
\begin{center}
\includegraphics[width=0.9\textwidth]{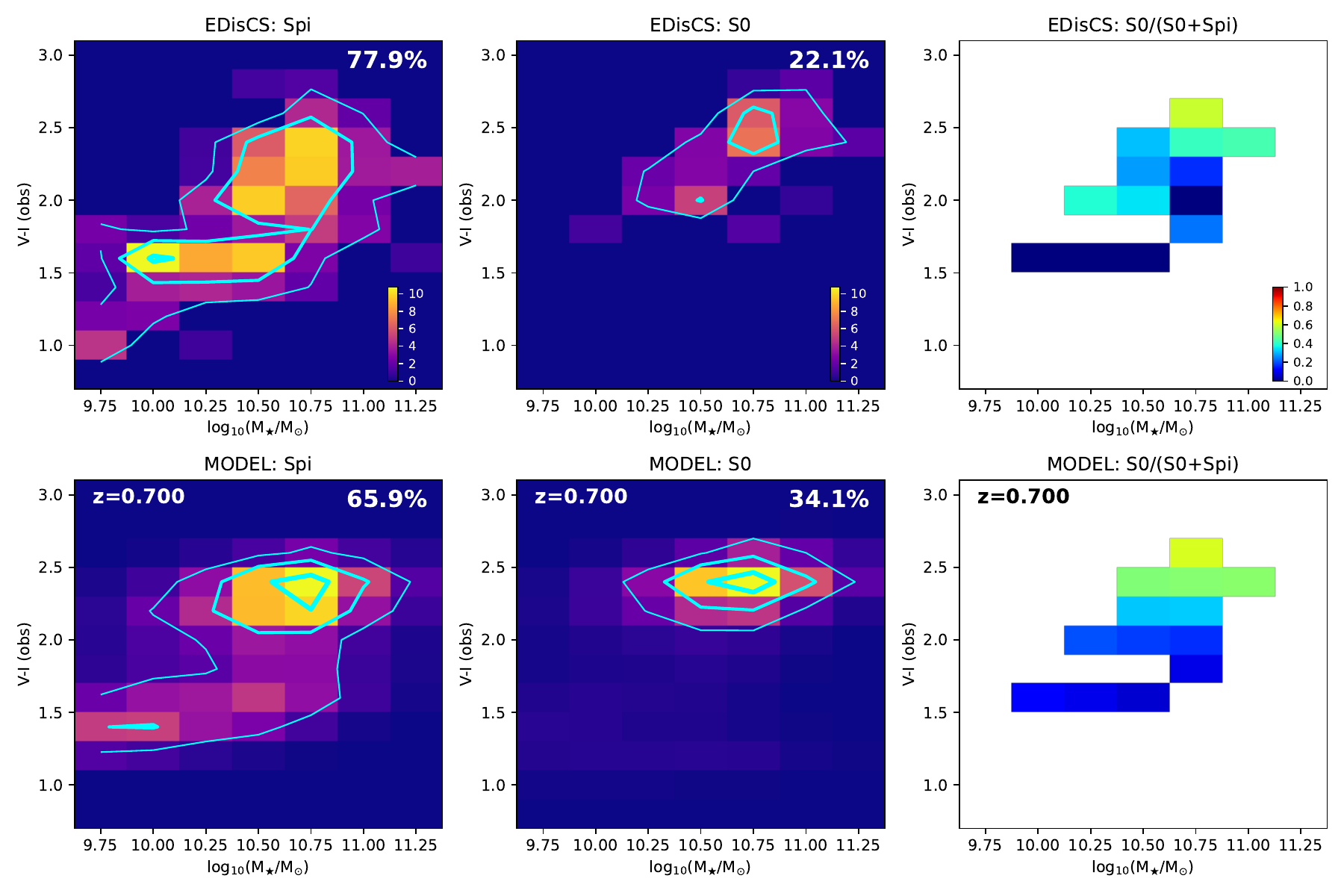}
\hrule
\includegraphics[width=0.9\textwidth]{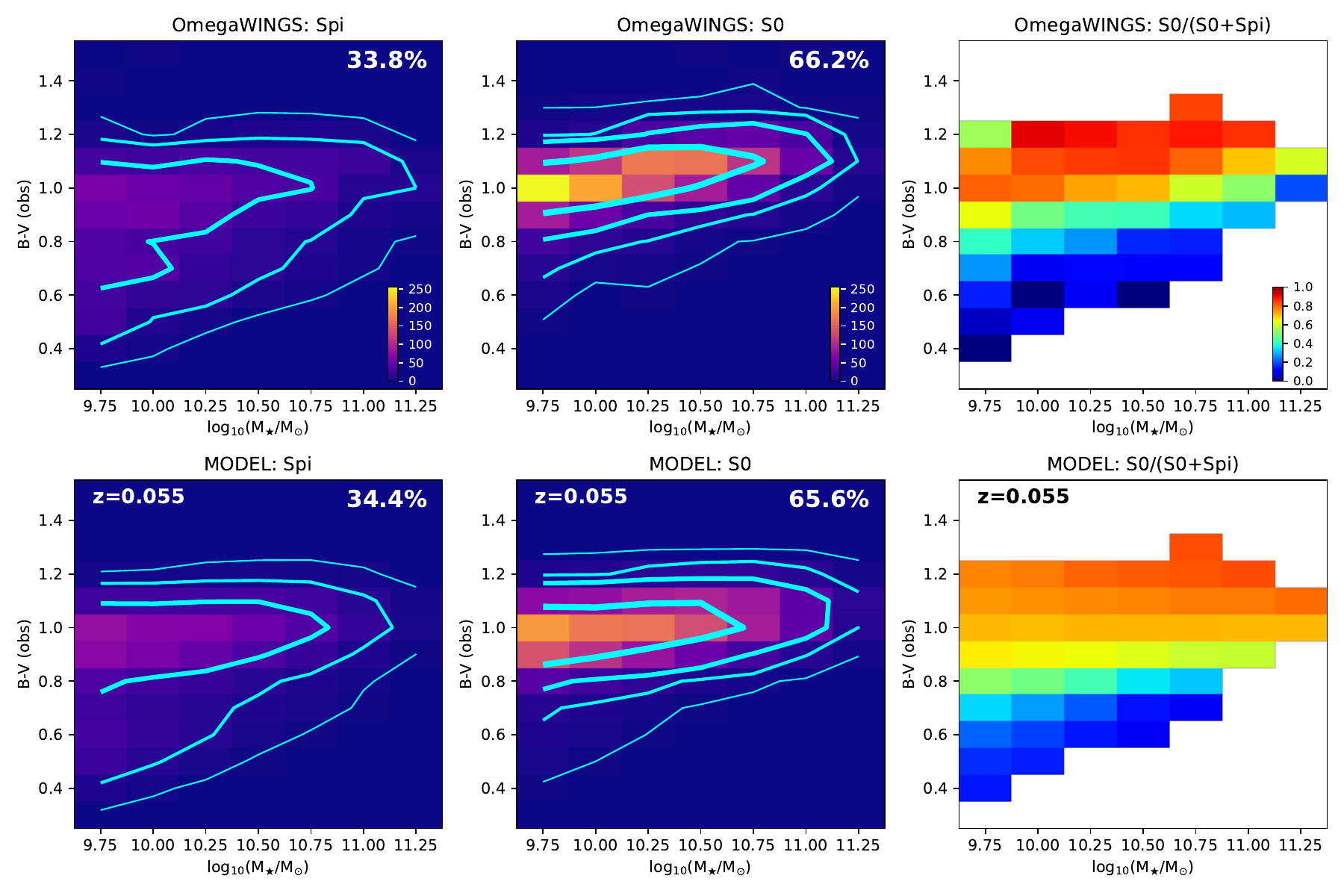}
\caption{Comparison of our best-fit cluster model with the EDisCS data at $z=0.7$ (top six panels), and with the OmegaWINGS data at $z=0.055$ (bottom six panels). The data and the model are shown in the first and second row of each panel set, respectively. 
Panels in the leftmost (central) column show the distribution of spirals (S0s) in a colour-$M_\star$ plane, where the colour is the observed-frame ($V-I$) for EDisCS and the ($B-V$) for OmegaWINGS. The colour palette indicate the number of galaxies per bin.
Cyan contours are drawn at $2$, $5$, $10$ and $15$ galaxies per bin in EDisCS and in the model at $z=0.7$, and at $2$, $10$, $30$ and $100$ galaxies per bin in OmegaWINGS and in the model at $z=0.055$. The total fraction of Spi and S0 galaxies with respect to the total disc (Spi+S0) population is indicated in the top-right corner of each panel.
Panels in the rightmost column show the S0 fraction as a function of colour and $M_\star$, for bins that have at least $5$ galaxies in the data.}
\label{f:hist2d_results}
\end{center}
\end{figure*}

Fig.\,\ref{f:hist2d_results} shows the comparison between the distribution of cluster galaxies in the color-$M_\star$ plane for the data and for our best-fit evolutionary model.
We first discuss the comparison with EDisCS at $z\!=\!0.7$, shown in the top-half of Fig.\,\ref{f:hist2d_results}.
The data show a dominance of Spi over the S0 population ($78\pm3\%$ vs $22\pm3\%$ of the total disc population), in agreement with what previously found for the same sample by \citet{Desai+07} and \citetalias{Vulcani+11}, and with other studies at similar redshifts \citep[e.g.][]{Dressler+94,Smail+97b,Fasano+00}.
The S0 population is built preferentially by high-$M_\star$ galaxies with $(V-I)\gtrsim2$, while Spi galaxies occupy also the bottom-left corner of the plane at lower $M_\star$ and bluer colours.
Interestingly, the data indicate a somewhat bimodal distribution of Spi galaxies, with an overdensity around $10.5\!<\!\log_{10}({M_\star}/M_\odot)\!<\!10.75$ and $2.0\!<\!(V-I)\!<\!2.5$ (which is similar to the region occupied by the S0s) and a separate bluer sequence at $\log_{10}({M_\star}/M_\odot)\!<\!10.5$ and $(B-I)\simeq1.5$.

Compared to the data, our best-fit model predicts a marginally more abundant S0 population at this redshift ($66\%$ Spi, $34\%$ S0), but it reproduces the observed trends quite well.
In particular, it does feature separate blue and red sequences for the Spi population, although it slightly overpredicts the number of red Spi hence providing a less populated blue sequence.
S0s in our model occupy a well defined region of the color-$M_\star$ plane, in excellent agreement with the data.
The rightmost panels in Fig.\,\ref{f:hist2d_results} show the S0 fraction (S0/(S0+Spi)) as a function of the galaxy colour and $M_\star$.
In this panel we have considered only bins containing at least $5$ disc galaxies, in order to have a minimum statistics.
As expected, the data show a trend for the S0 fraction to increase as a function of $M_\star$ and $(V-I)$, reaching a maximum value of $\sim0.5$. Our model reproduces this trend quite faithfully, reaching similar maximum fractions.

We now focus on the bottom-half of Fig.\,\ref{f:hist2d_results}, which shows the comparison with OmegaWINGS at $z=0.055$.
Compared with the high-$z$ population, the low-$z$ data tell a different story.
The majority of disc galaxies are S0s ($34\pm1\%$ Spi, $66\pm1\%$ S0).
The S0 population is distributed across the entire $M_\star$ range, with typical $(B-V)$ colours that are a slowly increasing function of $M_\star$: this trend is the well-known `ridge' of the red sequence \citep[e.g.][]{Baldry+04,Vulcani+22}.
The Spi population appears to be distributed similarly, but features a wider tail towards bluer colours for all $M_\star$ (the `blue cloud').
The combination of the observed trends leads to an S0 fraction that grows progressively from the blue cloud to the red sequence, reaching maximum values close to unity. 
The model provides an excellent match to all the observed trends: we clearly see the ridge of the red sequence and tail of the blue Spi building the blue cloud. 
The overall Spi and S0 fractions predicted by the model match the data almost perfectly.
However, there is a minor mismatch between the observed and predicted S0 fraction distribution, with the model featuring higher values within the blue cloud and lower values within the red sequence.
We argue that such mismatch could be amended with a minor ($<0.1$ mag) increase in the $(B-V)$ colour of the model S0 population, which would increase (decrease) the S0 fraction in the red sequence (blue cloud).
We also notice that the model predicts a red sequence that is slightly shallower than that observed.
We discuss this further in Section \ref{ss:model_variation}.

Overall, we find remarkable that, in spite of all the assumptions and simplifications we made, our model provides a cluster galaxy population that is realistic in terms of mass, colour and morphology distribution.
This is a necessary condition for the investigation of the physics of the cluster-driven evolution, which we examine below.

\subsection{RPS and Spi-to-S0 transformation timescales} \label{ss:timescales}
\begin{figure}
\begin{center}
\includegraphics[width=0.48\textwidth]{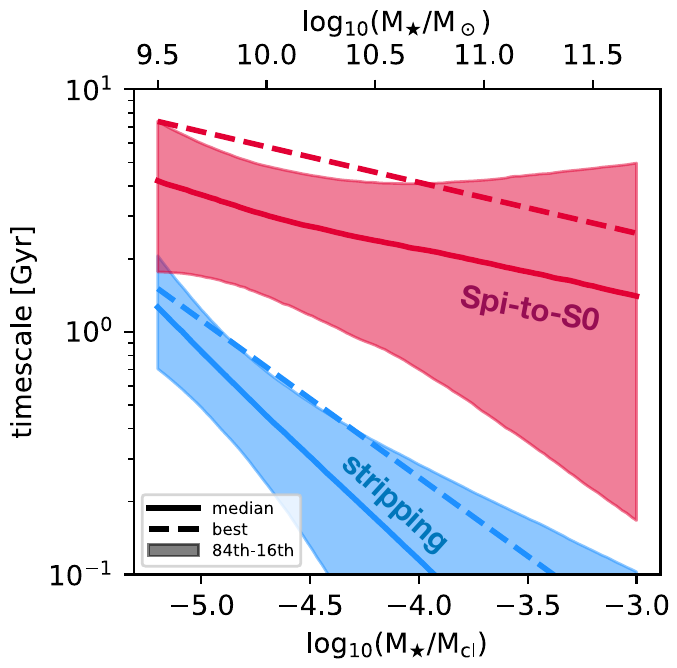}
\caption{Characteristic timescales for the ram-pressure stripping (blue) and for the cluster-driven morphological transformation (red) as a function of the galaxy-to-cluster mass ratio ($M_\star/M_{\rm cl}$) derived with our analysis (Equations \ref{eq:ts} and \ref{eq:tm}). 
The solid lines show the median trend, with the shaded regions showing the difference between the 16th and 84th percentiles.
The dashed lines show the best-fit model.
As a reference, in the horizontal axis on top we report the $M_\star$ scale for a fixed $\log_{10}(M_{\rm cl}/\msun)=14.7$.}
\label{f:timescale_results}
\end{center}
\end{figure}

The last two columns of Table \ref{t:param_Popesso} show the fiducial and the best-fit parameters of our model, while corner-plots showing the full posterior distributions are shown in Fig.\,\ref{f:cornerplot_Popesso}.
We find $p_{\rm max}\simeq0.8$, that is, there is a strong evidence for the morphological evolution of the cluster spiral population: in order to reproduce the data, it is required that $\sim80\%$ of spirals that join the cluster evolve into S0 on timescales that we discuss below.
We notice that the best-fit parameters of our model are within the quoted uncertainties of the fiducial ones, with the exceptions of $\tau_{\rm m}$ and $p_{\rm max}$. The latter takes of a value of $1$ for the model that maximises the likelihood.
We have attempted to re-fit our model by fixing $p_{\rm max}$ to $1$, finding fiducial values compatible with those reported in Table \ref{t:param_Popesso}.
The same applies for models that use the MS of \citet{Speagle+14}, instead of \citet{Popesso+23}, as reported in Table \ref{t:param_Speagle}.

We combine the posterior sampling provided by Nautilus with Equations (\ref{eq:ts}) and (\ref{eq:tm}) to investigate the `fiducial' relation between $t_{\rm s}$, $t_{\rm m}$ and $M_\star/M_{\rm cl}$ emerging from our model, and show our results in Fig.\,\ref{f:timescale_results}. 
We find that $t_{\rm m}\!>\!t_{\rm s}$ for all values of $M_\star/M_{\rm cl}$, implying that the morphological transformation occurs after ram-pressure stripping has removed the galaxy's gas reservoir, in agreement with our expectations.
The stripping timescale shows the strongest trend with the mass ratio, ranging from $\sim1\Gyr$ for $\log_{10}(M_\star/M_{\rm cl})\simeq-5$ and rapidly declining down to $\sim100\Myr$ at $\log_{10}(M_\star/M_{\rm cl})\simeq-4$.
This implies that, for a fixed cluster mass of $5\times10^{14}\msun$, galaxies with $M_\star\simeq5\times10^9\msun$ would have $t_{\rm s}\!=\!1\Gyr$, and galaxies with $M_\star\simeq5\times10^{10}\msun$ would have $t_{\rm s}\!=\!100\Myr$.
Overall, the range of $t_{\rm s}$ found here is largely compatible with those emerging from previous observational \citep[e.g.][]{Boselli+06,Jaffe+18} and theoretical \citep[e.g.][]{RoedigerBruggen07,Tonnesen+09,Marasco+16} studies, and supports RPS as the most efficient quenching mechanisms in clusters.
However, the anti-correlation between $t_{\rm s}$ and $M_\star/M_{\rm cl}$, caused by the negative $\alpha_{\rm s}$ found in our fitting, is counter-intuitive as one would expect most massive galaxies to be less influenced by the cluster environment, at a fixed $M_{\rm cl}$.
In Section \ref{ss:model_variation} we discuss the specific features in the data that drive this trend, and provide a possible interpretation in Section \ref{ss:anisotropy}.

Conversely, $t_{\rm m}$ shows a weaker correlation with $M_\star/M_{\rm cl}$, with typical values ranging from $1$ to $4\Gyr$, which qualitatively agree with the evolutionary timescales found by \citetalias{Marasco+23b}.
A more detailed comparison between our findings and those of \citetalias{Marasco+23b} is provided in Section \ref{ss:rps_evolution}.
We notice that there is a factor $\sim2$ difference between the fiducial and the best-fit $t_{\rm m}$ (solid and dashed red lines in Fig.\,\ref{f:timescale_results}).
This is likely driven by the (weak) degeneracy between $p_{\rm max}$ and $\tau_{\rm m}$, visible in the corner-plots of Fig.\,\ref{f:cornerplot_Popesso}, such that models with a more efficient morphological transformation (larger $p_{\rm max}$) but longer timescales (larger $\tau_{\rm m}$) are similar to those with a less efficient transformation but shorter timescales.

\subsection{Evolution of the morphological mix since $z\simeq0.6$} \label{ss:evolution_morphmix}
\begin{figure}
\begin{center}
\includegraphics[width=0.45\textwidth]{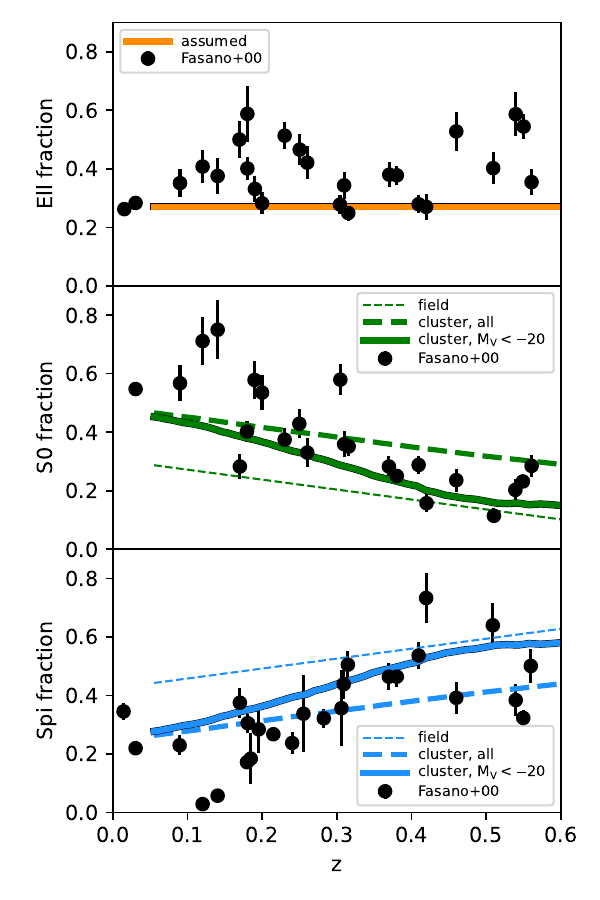}
\caption{Evolution of the morphological fraction of cluster galaxies for $z<0.6$. The black circles with error-bars show the measurements of \citet{Fasano+00}, complete down to $M_{\rm V}\!<\!-20$.
The thick solid curves show the predictions from our best-fit model using the same photometric completeness, while the thick dashed curves show the predictions using all galaxies available.  
The thin dashed curves show the assumed field fractions. 
As discussed in Section \ref{sss:field_morphology}, the Ell fraction in the model is set to a constant $27\%$ by construction.}
\label{f:fasano_comparison}
\end{center}
\end{figure}
We now focus on the evolution of the morphological mix in galaxy clusters.
A key observational study in this context is that of \citet{Fasano+00}, who investigated the morphological content of clusters over the redshift range $0\!<\!z\!<\!0.6$. 
They used an heterogeneous sample made of $25$ clusters at intermediate redshifts observed either with HST/WFPC2 or with high-quality ground-based imaging, complemented by a local comparison set of $22$ nearby clusters drawn from \citet{Dressler+80} catalogue. 
All galaxies were morphologically classified into Ell, S0s, and Spi, and particular care was devoted to ensuring consistency across the different datasets via cross-checks between HST and ground-based classifications.

In Fig.\,\ref{f:fasano_comparison} we compare the evolution of the morphological fractions found by \citet{Fasano+00}, complete for galaxies down to $M_{\rm V}=-20$, with the prediction of our best fit model.
Predictions that use the same magnitude completeness of the data (thick solid lines) are shown separately from those that include all galaxies with $M_\star\!>\!10^{9}\msun$ available in the model, in order to highlight the importance of the photometric selection.
The predictions for the model obtained with the fiducial parameters are very similar and are not shown here for brevity.
The data show a clear trend with redshift, featuring a steady increase of the S0 fraction at the expenses of the Spi, with the fraction of Ell that remains approximately constant.
Our model follows very similar trends, especially when the magnitude cut is applied, but with two main differences. 
First, our Ell fraction is fixed by construction to $27\%$ (Section \ref{sss:field_morphology}), which is valid for the local field population but is somewhat a lower limit for the cluster population.
Second, the Spi-to-S0 evolution predicted by the model is slightly milder than that shown by the data, which, compared to the model, feature a higher (lower) S0 (Spi) fraction in local galaxy clusters.
However, as our model reproduces with remarkable accuracy the Spi and S0 fractions of OmegaWINGS clusters, this issue appears to be associated either with limitations in the dataset chosen to constrain the model, which cover only the initial and final stages of the observed evolution, or with a substantial scatter in the cluster morphological mix, such that the pronounced trend visible in the data of \citet{Fasano+00} is caused by stochastic fluctuations in cluster sample.

In general, our results agree with those of \citetalias{Vulcani+11}, who found that the EDisCS-to-OmegaWINGS evolution is driven by the combined effect of infall, star formation, and morphological evolution of late-types into early-types.
However, our analysis would benefit from tighter observational constrains from novel observations of field and cluster galaxies at various redshift.
This will be possible in the near future thanks to the CHileAN Cluster galaxy Evolution Survey \citep[CHANCES;][]{Haines+23,Sifon+25}, a spectroscopic survey of galaxies in clusters and their surroundings up to $z\simeq0.45$ with the 4MOST instrument at the VISTA telescope \citep{4MOST}, and to new observational facilities such as the Euclid telescope \citep{Euclid} or the Vera Rubin observatory \citep{LSST}.

\section{Discussion}\label{s:discussion}
\subsection{Model limitations}\label{ss:model_limitations}
Our models are designed to describe the complex interplay between galaxies and the cluster environment in a simplified (yet flexible) way, and with an affordable computational cost.
Apart from the simplified treatment of the environmental mechanisms, perhaps their main limitation is the treatment of galaxy clusters as unique, homogeneous entities, within which galaxies that are accreted from the field (which is also considered homogeneous) evolve according to parametrised recipes.
We do not treat effects that are associated with the cluster-centric distance or with the local density, although these are observed to have an impact on the properties of galaxies in clusters \citep[e.g.][]{Perez-Millan+23} and in groups \citep[e.g.][]{George+13}.
This simplification bypasses detailed treatments for the ICM properties, allowing us to explore a broader parameter space more efficiently.
However, it prevents us from investigating the smooth transition of galaxy properties from the field to the cluster environments, which occurs in a transition region that extends beyond the cluster virial radius \citep[e.g.][]{Vulcani+23}.

A related concern is the lack of environment classes aside from clusters and the field, whereas observations and simulations show that $30-50\%$ of galaxies reaching clusters today arrive as part of accreted groups rather than individually \citep[e.g.][]{McGee+09, DeLucia+12, Benavides+20}.
Although we do not discuss specifically the role of pre-processing \citep[e.g.][]{ZabludoffMulchaey98,Fujita04,Sharonova+25}, we stress that the field morphologies adopted at $z\!=\!0.055$ (Section \ref{sss:field_morphology}) are compatible with the actual morphological mix observed around $2\times R_{200}$ in OmegaWINGS clusters \citep[Fig.\,2 of][]{Vulcani+23}. 
This indicates that, in our model, the properties of the galaxy population infalling onto local clusters are realistic.
We discuss further experiments with variations in the field morphology mix in Section \ref{ss:field_extreme}.

Previous studies that modelled the evolution of satellite galaxies in clusters and groups have introduced the so called `delayed-than-rapid' quenching scenario, where the host environment begins to affect galaxies only past a given delay time $t_{\rm del}$ after their infall \citep[e.g.][]{Wetzel+13,OmanHudson16,Fossati+17}.
We have verified that the introduction of a delay time parameter in our model - a unique value valid both for the RPS and for the morphology transformation - does not improve the overall quality of the fit, leading to very short $t_{\rm del}$ (best-fit value of $0.2\Gyr$, fiducial value of $0.6\pm0.4\Gyr$) and practically the same likelihood and characteristic timescales of our previous model.
A $t_{\rm del}$ of a few Gyr helps to increase the overall Spi fraction at $z=0.7$ due to the delayed transformation, but leads to too many blue galaxies at that redshift due to their delayed quenching.
Similarly, it helps to mitigate the excess of low-mass red spirals at $z=0.055$ (see Section \ref{ss:model_variation}), but also produces too many spirals at that redshift.
These issues could be alleviated by invoking separate delay times for the morphological and the colour evolutions, or a redshift-dependent $t_{\rm del}$.
However, exploring these additional parameters goes beyond the purpose of this study.

Finally, our models account only for the `negative' effect of ram pressure on star formation, but both observations \citep{Merluzzi+13,Vulcani+18a,Vulcani+20} and theoretical models \citep{FujitaNagashima99,Tonnesen+09,Bekki+14} indicate that, prior to complete gas removal, the star formation activity can be enhanced due to gas compression \citep[e.g.][]{Moretti+20b,Moretti+20a}.
Observations suggest that this enhancement can be particularly strong in gas-rich dwarf galaxies \citep{Grishin+21}, and could contribute to populate the sequence of low-$M_\star$ blue spirals which in our model, especially at $z\!=\!0.7$, is somewhat deficient.

\subsection{Variations in the model parameters}\label{ss:model_variation}
\begin{figure*}
\begin{center}
\includegraphics[width=0.95\textwidth]{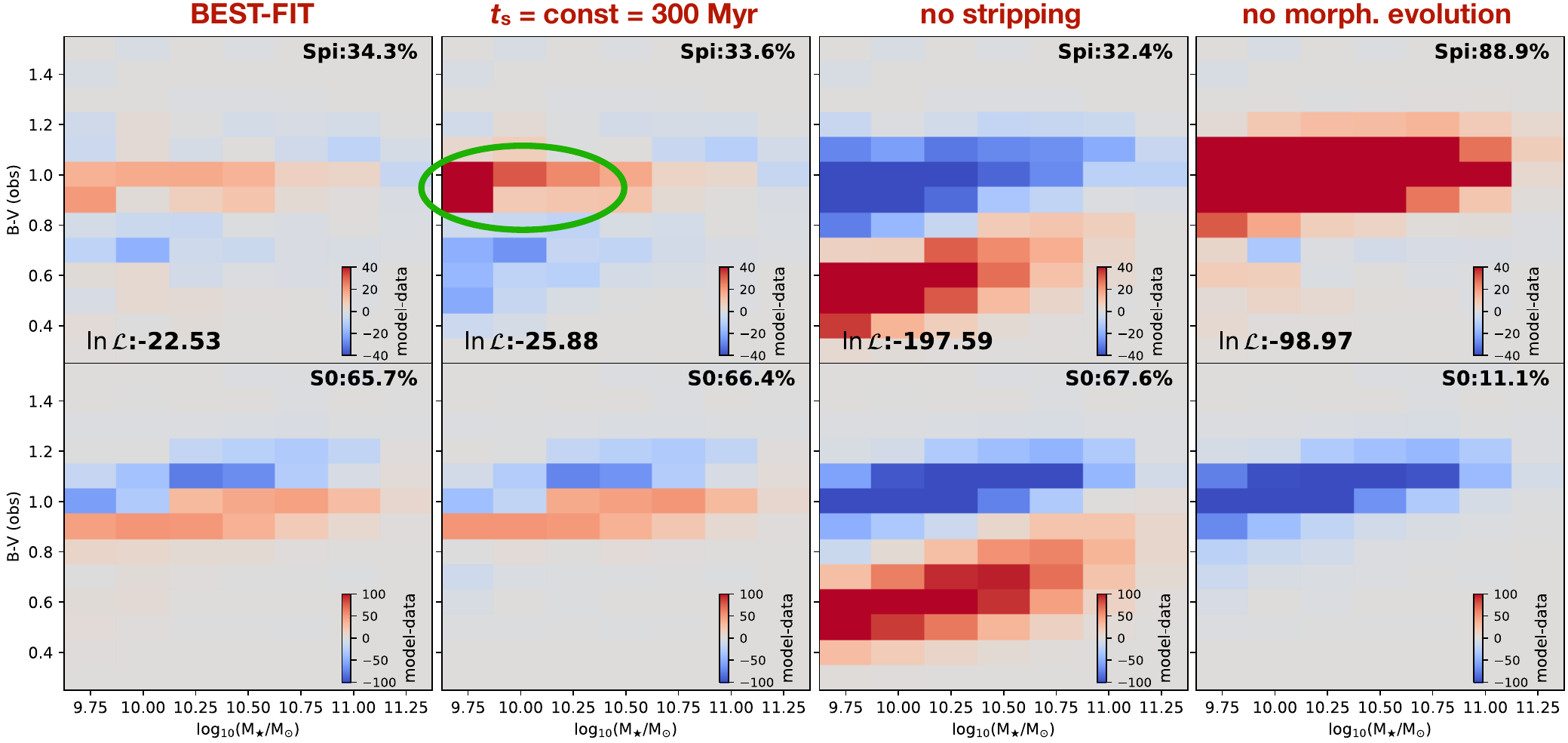}
\caption{Comparison between the residuals (model-data) in the $(B-V)$-$M_\star$ plane of four different models: our best-fit model (\emph{first column}), a model with constant $t_{\rm s}$ of $300\Myr$ (\emph{second column}), a model without RPS from the ICM (\emph{third column}), and a model without morphological evolution (\emph{fourth column}). Red (blue) colours indicate an excess (a deficiency) of galaxies in the model compared to the data. 
The top and bottom rows show the Spi and S0 populations, respectively. 
The green ellipse highlights the excess of low-mass red spirals in the model with constant $t_{\rm s}$.
For each model we also list the value of the log likelihood (Equation \ref{eq:loglk}). For brevity, we only show the results at the redshift of the OmegaWINGS clusters.}
\label{f:residuals}
\end{center}
\end{figure*}
To better assess the performance of our approach, we now compare our best-fit model with three variant models where we have modified individual values of the best-fit parameters.
In the first variant model we have fixed the stripping timescale to $300\Myr$ and removed any trend with mass ($\alpha_{\rm s}\!=\!0$).
In the second, we have increased $t_{\rm s}$ to an arbitrarily large value, effectively removing the cluster-driven quenching.
In the third, we have removed the cluster-driven morphological transformation by imposing $p_{\rm max}\!=\!0$.

The results of these experiments are presented in Fig.\,\ref{f:residuals}, where we compare the residuals (model-data) in the $(B-V)$-$M_\star$ plane at the redshift of OmegaWINGS for the four different models. 
Clearly, the best-fit model provides the smallest residuals and the highest likelihood value.
As already mention in Section \ref{ss:data_vs_model}, our model appears to slightly underestimate the colour of the S0 population, leading to the offset visible in the bottom-left panel of Fig.\,\ref{f:residuals}.
It appears that this limitation is intrinsic to our scheme and we cannot amend it with any choice of the model parameters. 
Redder colours in our model could be obtained by increasing the metallicity of the quenched galaxies, or by using different SPS models.
Interestingly, the model with $t_{\rm s}\!=\!300\Myr$ appears almost identical to our best-fit one, with a single exception: it overestimates the number of red spirals in the low-$M_\star$ regime. 
The red spiral excess, highlighted with a green ellipse in Fig.\,\ref{f:residuals}, is very severe in the lowest bin of $M_\star$ but remains visible up to $\sim 10^{10.3}\msun$, well above the mass completeness limit in OmegaWINGS.
This is a critical feature that pushes the model towards $\alpha_{\rm s}\!<\!0$: low-$M_\star$ spirals must have longer stripping timescales (as we have seen in Section \ref{ss:timescales}) in order to avoid overpopulating the red sequence.
This result holds even when we exclude the lowest $M_\star$ bin from our fit.

The models that exclude either RPS or the morphological transformation provide significantly worse residuals and much lower likelihood values.
Excluding RPS leads to overpopulate the blue cloud with both spirals and S0s (and, consequently, to underpopulate the red sequence) due to the fact that, while the transformation from Spi to S0 proceeds regularly, galaxies remain on the MS of star formation unless they have joined the cluster as S0 to begin with.
Finally, excluding the morphological evolution leads to drastically overestimate the number of spirals, and to underestimate the number of S0s, everywhere in the color-$M_\star$ plane.
We stress that this result holds even if we had fixed the field morphological fractions at all redshifts to the values of \citet{Calvi+12}, valid at $z\!\simeq\!0$, in order to minimise the efficiency of the morphological evolution.
Unless the field morphology fractions are significantly different from those adopted here (see Section \ref{ss:field_extreme}), cluster-driven morphological transformation is necessary to explain the data.

\subsection{Is RPS the main driver of the morphological evolution?}\label{ss:rps_evolution}
\begin{figure}
\begin{center}
\includegraphics[width=0.48\textwidth]{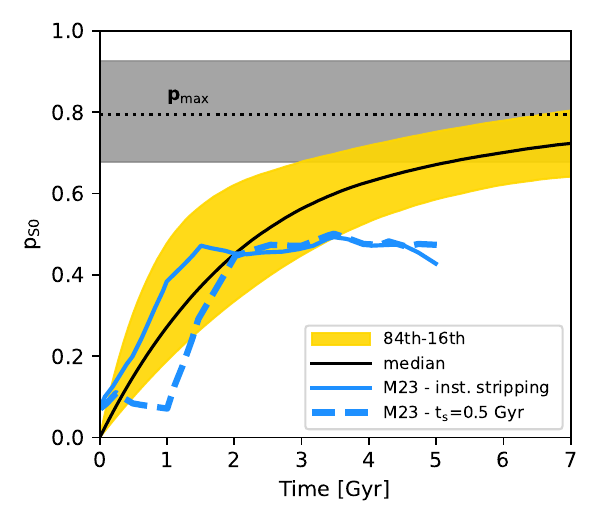}
\caption{Spiral-to-S0 transformation probability as a function of the time from the galaxy's entry into the cluster. The blue curves show the results of \citetalias{Marasco+23b} for the spectrophotometric evolution of a sample of $15$ field galaxies with typical $\log_{10}(M_\star/M_\odot)\!\simeq\!10.5$, assuming either instantaneous stripping (solid) or $t_{\rm s}\!=\!0.5\Gyr$ (dashed). The solid black curve shows our fiducial model for $\log_{10}(M_\star/M_\odot)\!=\!10.5$ and $\log_{10}(M_{\rm cl}/M_\odot)\!=\!14.7$, with yellow shaded region showing its scatter given by the difference between the 84th and 16th percentiles. The dotted horizontal line with the grey band on top shows our fiducial $p_{\rm max}$ and its scatter.}
\label{f:pS0_M23}
\end{center}
\end{figure}
Although the modelling adopted for the morphological transformation of cluster galaxies described in Section \ref{sss:model_morphevo} is inspired by the results of \citetalias{Marasco+23b}, its mathematical implementation is rather general, as the transformation channel is not explicitly specified. 
The channel can be inferred a posteriori by comparing the resulting transformation timescales with those appropriate for specific channels.
However, the right-hand panel of Fig.\,\ref{f:timescale_results} shows quite some scatter in $t_{\rm m}$, with values ranging from less than $1$ to several Gyr, which complicates the interpretation of our results.

We show a more detailed comparison with the spectrophotometric evolution results of \citetalias{Marasco+23b} in Fig.\,\ref{f:pS0_M23}. 
Here, we have focussed on the evolution of the so called `control field' (CF) galaxies \citep{Vulcani+18a}, a subsample of GASP made of $15$ field galaxies with typical $\log_{10}(M_\star/M_\odot)\!\simeq\!10.5$, which provided the most reliable evolutionary trends due to their regularity and absence of substructures (see Section 3.3 in \citetalias{Marasco+23b} for additional details).
In Fig.\,\ref{f:pS0_M23} we compare the evolution of $p_{\rm S0}$ determined for these galaxies by \citetalias{Marasco+23b} (followed up to $5\Gyr$ from the onset of the RPS) with that inferred via Equation (\ref{eq:pS0}) with our model for $\log_{10}(M_\star/M_\odot)\!=\!10.5$ and $\log_{10}(M_{\rm cl}/M_\odot)\!=\!14.7$.
For the CF galaxies of \citetalias{Marasco+23b} we show the trends for two stripping scenarios, instantaneous and gradual stripping with $t_{\rm s}=0.5\Gyr$, which brackets the $t_{\rm s}$ range predicted by our model for the $M_\star$ and $M_{\rm cl}$ considered (left-hand panel of Fig.\,\ref{f:timescale_results}).

Clearly, the trends predicted by our evolutionary model appears to be consistent with those of the spectrophotometric model for the initial $\sim2\Gyr$.
Beyond that point, the morphology of galaxies ceases to be affected by the ageing of their stellar population: their spectrophotometric evolution is completed, reaching an asymptotic $p_{\rm max}\simeq0.5$.
In our evolutionary model, instead, $p_{\rm S0}$ keeps on growing with time, flattening to a slightly larger $p_{\rm max}\simeq0.8$.
One the one hand, this may point to additional evolutionary channels that keep shaping the morphology of cluster galaxies on longer timescales, such as secular processes \citep{SellwoodCarlberg84, Fujii+11} or, for the low-$M_\star$ regime, harassment \citep{Moore+96, Smith+10}.
On the other hand, we stress that the trends resulting from the analysis of \citetalias{Marasco+23b} are very conservative due to their neglect of stellar kinematics, and in particular of disc rotation, which would smear out the substructures present the galaxy and boost the transition to the S0 type.
Additionally, the blue curves in Fig.\,\ref{f:pS0_M23} represent mean trends valid for the whole CF population, but individual galaxies can feature larger $p_{\rm max}$.

These considerations do not allow us to identify a unique transformation channel, but we can safely state that our results are consistent with the spectrophotometric ageing being a key channel for the spiral-to-S0 transformation observed in cluster galaxies at $z\lesssim1$.

\subsection{Exploring `extreme' field morphologies}\label{ss:field_extreme}
\begin{figure}
\begin{center}
\includegraphics[width=0.45\textwidth]{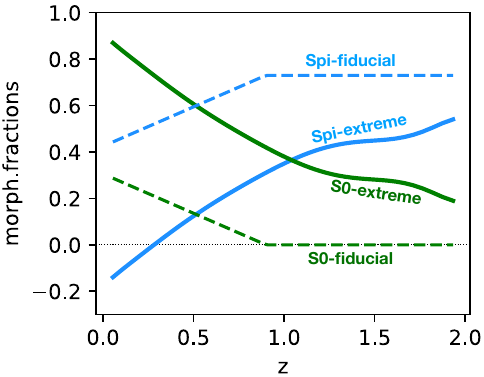}
\caption{Comparison between the field morphological fractions used in our main analysis (dashed curves) and those inferred for an extreme model where the field is fully responsible for the morphological evolution of cluster galaxies (solid line). In both cases the Ell fraction is fixed to $0.27$ at all redshifts.}
\label{f:field_extreme}
\end{center}
\end{figure}

One of the main ingredient of our model is the morphological fraction in the field, which we assumed to vary with redshift following the trends shown in the bottom panel of Fig.\,\ref{f:halo_growth}.
As these trends are poorly constrained observationally, alternative scenarios deserve to be discussed, at least from a qualitative standpoint.

We consider here two extreme scenarios.
In the first one, the field morphological mix does not vary with $z$ and stays constant to the local values found by \citet{Calvi+12}. 
We have refit the data under such conditions, finding that morphological transformation is still required to reproduce the OmegaWINGS data, and with $t_{\rm m}$ compatible with those of our fiducial model. 
However, such model largely overestimates the S0 fraction in EDisCS.
Trivially, this is caused by the Spi fraction in EDisCS clusters being larger than that in the local field, which itself is an indirect evidence for morphological evolution of the field population.

In the second scenario, we assume that the cluster-driven Spi-to-S0 evolution is suppressed, and it is the field morphology that is adjusted to match the cluster data.
In practice, in this scenario, we consider pre-processing as the only responsible for the observed morphology mix in clusters.
A simple way to investigate this scenario is to `reverse engineering' our best-fit model: at all redshifts, we determine the field fractions that would be required to match the morphological mix of the best-fit cluster model in the absence of internal evolution.
The fractions inferred in this way are shown in Fig.\,\ref{f:field_extreme}. 
Clearly, matching the cluster data requires a field dominated by S0s for $z\!<\!1$, and with an S0 fraction that steadily increases at later times, in strike contrast with the morphologies observed in the local field and at the periphery of OmegaWINGS clusters \citep{Vulcani+23}.
Paradoxically, this approach demands negative spiral fractions at later times, caused by the fact that, in the model, the total number of spiral cluster galaxies decrease with time as the morphological transformation proceeds at a rate faster than that at which new spirals are injected from the field.
These findings exclude preprocessing as the main responsible for the morphology mix observed in local clusters, corroborates our original choice for the field morphology fractions, and strengthen the overall results of our analysis.

\subsection{Stripping and orbit anisotropy}\label{ss:anisotropy}
In Section \ref{ss:timescales} we have shown that, in our models, $t_{\rm s}$ and $M_\star/M_{\rm cl}$ anti-correlate: for a fixed cluster mass, the stripping proceeds more efficiently in high-$M_\star$ galaxies, which is unexpected.
One possible interpretation of this puzzling result is that galaxies with different $M_\star$ join the cluster following substantially different trajectories, with the least (most) massive systems having a stronger tangential (radial) component in their initial velocity.

Strongly radial orbits lead galaxies closer to the pericenter and with a larger speed, maximising the RPS efficiency. 
We notice that a high RP does not necessarily imply a large occurrence of jellyfish systems, given that low-mass galaxies (which are the most numerous) will be stripped on very short timescales, limiting the possibility to observe the mechanism in action \citep[e.g.][]{Marasco+16}.
Observations of jellyfish galaxies in different mass regimes appear to support this, indicating preferentially radial (tangential) orbits for high- (low-) mass jellyfishes \citep{Grishin+21,Biviano+24}.

Several studies have investigated the orbit anisotropy in galaxy clusters as a function of galaxy mass, morphology and clustercentric distance, using either dynamical modelling of the observed cluster kinematics \citep[e.g.][]{Adami+98,Annunziatella+16,Biviano+21} or large-scale cosmological simulations \citep[e.g.][]{Lotz+19}.
The general picture emerging from these studies is that the anisotropy parameter $\beta$ declines from positive values (indicating preferentially radial orbits) in the cluster outskirts, where star-forming galaxies reside, to negative values (tangential orbits) in the cluster core, which is dominated by passive, massive systems.
This supports a scenario where low-mass, star-forming systems joining the cluster from the field with $\beta\!>\!0$ are tidally destroyed, biasing the surviving population within the core towards $\beta\!<\!0$.

While the results discussed above are valid for the cluster population as a whole, our findings concern the infalling population alone.
The existence of a correlation between $M_\star$ and $\beta$ for the galaxies that join the cluster from the field is yet to be proven, and it may be challenging to probe it observationally.
Instead, it should be straightforward to investigate it using cosmological hydrodynamical simulations.
However, this goes beyond the purpose of our study.

\section{Conclusions}\label{s:conclusion}
Galaxy clusters are formidable laboratories to study the effect of the environment on the evolution of their members. 
Compelling observational evidence indicate a rapid evolution in the properties of the disc galaxy population in clusters over the past $\simeq7\Gyr$, with passive, lenticular (S0) systems dominating in local clusters, and more actively star forming spirals (Spi) dominating the population at $z\simeq0.7$ and beyond.
This evolution does not match that observed in the field, implying an active role of the cluster environment in driving the observed trends \citep[e.g.][]{Guglielmo+15}.

In this study, we have investigated whether the observed evolution can be explained by simple cluster evolutionary models in the $\Lambda$CDM framework, where the main mechanisms driving the evolution are ram pressure stripping (RPS) from the intra-cluster medium and a gradual transformation of cluster spirals into S0s due to an (initially) unspecified process.
We assume that the main model parameters, the stripping timescale $t_{\rm s}$ and the Spi-to-S0 transformation timescale $t_{\rm m}$, are independent power-law functions of the galaxy-to-cluster mass ratio $M_\star/M_{\rm cl}$, and constrain their values using the observed spiral and S0 distributions in the color-$M_\star$ plane from the EDisCS survey at $z\simeq0.7$ and the OmegaWINGS survey at $z\simeq0.055$.
Our results can be summarised as follows:
\begin{itemize}
    \item Our best-fit model reproduces the observed distribution of spirals and S0 cluster galaxies in the colour-$M_\star$ plane very well, especially at low redshift (Fig.\,\ref{f:hist2d_results}), where the model has vastly lost memory of its initial conditions. However, compared to the data, our model slightly overestimates the S0 fraction at $z=0.7$, and underestimate the $(B-V)$ colour of the S0 population at $z=0.055$ by $\sim0.1$ dex (bottom-left panel of Fig.\,\ref{f:residuals}).
    \item We find typical $t_{\rm s}$ between $100\Myr$ and $1\Gyr$, in agreement with previous estimates \citep{Tonnesen+09,Jaffe+18,Salinas+24}. 
    Typical $t_{\rm m}$ are instead of a few Gyr, albeit with a large scatter (Fig.\,\ref{f:timescale_results}).
    We interpret the fact that $t_{\rm m}\!>\!t_{\rm s}$ in terms of cause and effect: the morphological transformation follows the quenching of star formation caused by RPS.
    \item The morphological evolution inferred from our model is compatible with that resulting from the simple ageing of the stellar populations in outside-in quenched galaxies, proposed by \citetalias{Marasco+23b}.
    Albeit we cannot exclude that secular evolution plays a role in shaping galaxy morphology on longer timescale, our results confirm spectrophotometric ageing as a key channel for the Spi-to-S0 transition in galaxy clusters.
\end{itemize}

An unexpected and somewhat controversial finding is the strong anti-correlation between $t_{\rm s}$ and $M_\star/M_{\rm cl}$, driven by the relatively blue colour of low-$M_\star$ spirals in the OmegaWINGS clusters: assuming a constant $t_{\rm s}$ at all masses leads to overestimate the number of low-$M_\star$ red spirals compared to the data (Fig.\,\ref{f:residuals}, second panel in the top row).
One possible interpretation is provided in terms of orbit anisotropy, discussed in Section \ref{ss:anisotropy}, but a detailed comparison with large-scale cosmological simulations is required to validate this scenario. 

Our analysis is based on detailed information on the cluster galaxy populations available in two specific redshift bins.
We expect forthcoming observational datasets from the CHANCES spectroscopic survey and from new facilities, such as the Euclid telescope and the Vera Rubin Observatory, to provide tighter constraints to our model over a larger redshift range, significantly boosting our understating of the role played by dense environments in the formation and evolution of galaxies.


\begin{acknowledgements}
The authors thank an anonymous referee for a thoughtful and constructive report.
AM acknowledges funding from the INAF Mini Grant 2023 program `The quest for gas accretion: modelling the dynamics of extra-planar gas in nearby galaxies'. BV acknowledges support from the INAF GO grant 2023 ``Identifying ram pressure induced unwinding arms in cluster spirals'' (P.I. Vulcani).
\end{acknowledgements}

\bibliographystyle{aa} 
\bibliography{cluster_evo} 

\begin{appendix}
\section{Relation between star formation history and main sequence} \label{a:SFH_MS}

\begin{figure*}
\begin{center}
\includegraphics[width=0.95\textwidth]{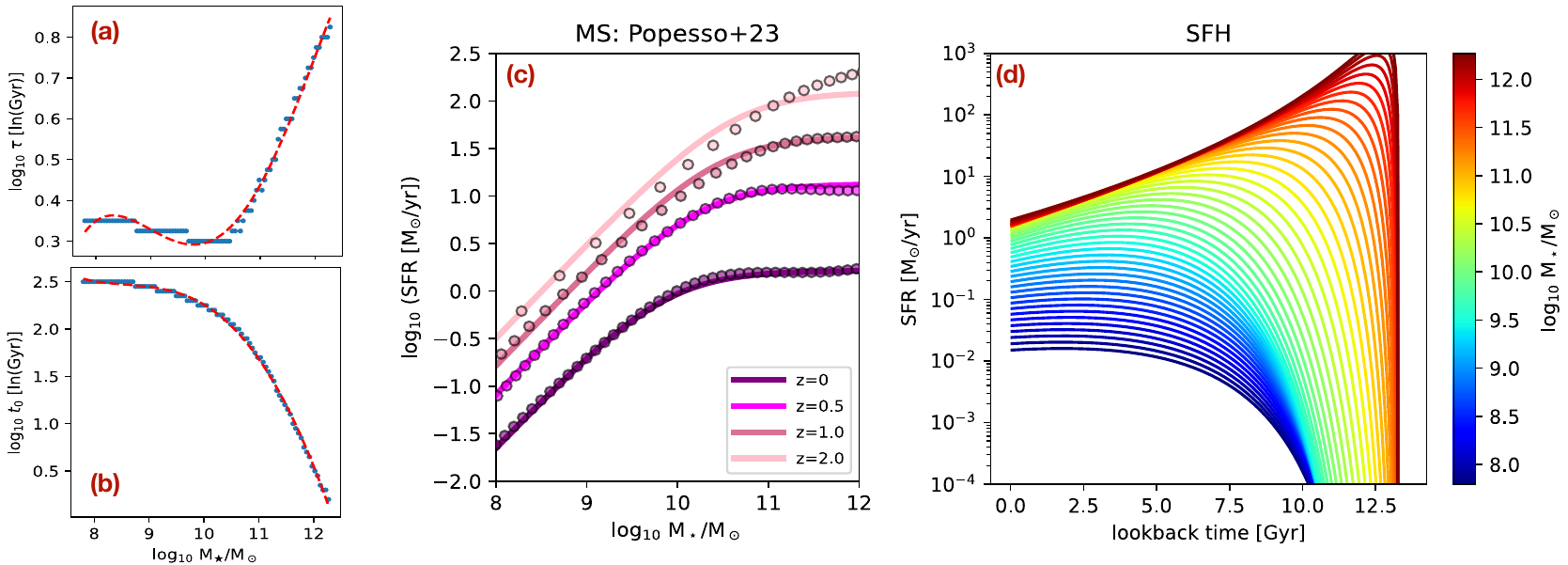}
\caption{\emph{Panels (a) and (b)}: relation between $\tau$ ($a$) and $t_0$ ($b$), the two parameters of our SFH model (Equation\,\ref{eq:lognorm}), and the galaxy $M_\star$ at $z=0$, constrained by the \citet{Popesso+23} main sequence (MS) of star formation. The blue dots show the models in our library that minimise the distance to the MS at various $z$. The red-dashed curves show our regularisation using 5th degree polynomials. \emph{Panel (c)}: SFR-$M_\star$ relations at four different redshifts. The solid curves show the MS of \citet{Popesso+23}, the filled circles show the sequence emerging from our fiducial SFH library. \emph{Panel (d)}: our fiducial SFH library adopted for MS galaxies, color-coded by the galaxy $M_\star$ at $z\!=\!0$. Our models support the downsizing scenario of $M_\star$ build-up.}
\label{f:tau_t0_mass_Popesso}
\end{center}
\end{figure*}

In this Appendix we provide details on the SFH models adopted for our main-sequence field galaxies at the various redshifts.
Following \citet{Gladders+13}, we use log-normal SFH models that are described as:
\begin{equation}\label{eq:lognorm}
    {\rm SFR}(t,t_0,\tau) = \frac{1}{t\sqrt{2\pi\tau^2}}\,\exp\left[-\frac{(\ln t-t_0)^2}{2\tau^2}\right]
\end{equation}
where $t$ is the elapsed time since the big bang, $t_0$ is the logarithmic time delay, and $\tau$ sets the rise and decay timescale.

Our goal is to find optimised relations between the two parameters of Equation\,(\ref{eq:lognorm}), $t_0$ and $\tau$, and the $M_\star$ of a $z\!=\!0$ galaxy, so that such galaxy stays (approximately) on the MS of \citet{Popesso+23} up to $z\simeq2$.
This is achieved numerically with the following approach.
First, we build an initial, large ($\sim2\times10^5$) family of model SFHs with varying values of $t_0$, $\tau$, and $M_{{\rm f},0}$, the total gas mass converted into stars by $z=0$ given by the integral of Equation\,(\ref{eq:lognorm}).
For each model we store also total gas mass converted into stars at any $z$, $M_{{\rm f},z}$.
Next, we convert our $M_{{\rm f},z}$ to $M_\star$ estimates, which exclude the gas lost by stellar winds and supernovae (but include stellar remnants), using the returned gas fractions provided by the \citet{BruzualCharlot03} SPS model (2016 version) available in \bagpipes.
These depend mainly on the galaxy SFH and only to a minor extent on the stellar metallicity, which is assumed to be Solar for the purpose of this calculation.

We then exclude from our family all models that provide $M_\star$ and SFRs incompatible with the MS. 
This is done by evaluating the typical `distance' between our models and the MS at four representative redshifts ($0$, $0.5$, $1$, $2$), and then excluding models with distance larger than a fixed threshold.
This filters out the vast majority of models, leaving only those that are compatible with the MS.
As we show in panels (a) and (b) of Fig\,(\ref{f:tau_t0_mass_Popesso}), the surviving models show clear correlations between $M_\star$ and the parameters of Equation\,(\ref{eq:lognorm}).
We regularise these correlation using fifth-degree polynomials, which optimally describe the observed trends.
Finally, the polynomial fits are used to build our fiducial SFH library for MS galaxies, which we utilise in our cluster evolutionary model.

Panel (c) of Fig.\,(\ref{f:tau_t0_mass_Popesso}) shows the comparison between the MS of \citet{Popesso+23} and those predicted by our fiducial SFH library.
Clearly, these can reproduce very well the `bending' of the MS visible at low $z$ while still providing a good match to the high-$z$ MS, although their accuracy gets somewhat worse around $z\simeq2$. 
The resulting sequence of SFHs, shown in panel (d) of Fig.\,(\ref{f:tau_t0_mass_Popesso}), presents clean distinction between low- and high-$M_\star$ galaxies, with the former showing gently rising SFHs, while the latter feature a sharp peak in their SFR at earlier epochs, followed by a gradual decline.
These trends agree with the so-called `downsizing' scenario \citep[e.g.][]{Cowie+96}, where more massive galaxies have assembled their $M_\star$ faster and at earlier epochs than the low-mass ones.

The same technique adopted here can be applied to any MS of star formation, provided that its evolution with $z$ is known.
We have applied our method to the MS of \citet{Speagle+14}, finding analogous correlations between $M_\star$, $t_0$ and $\tau$, and qualitatively similar SFHs supporting the downsizing scenario.

\section{Supplementary material} \label{a:supplementary}
\begin{figure*}
\begin{center}
\includegraphics[width=0.75\textwidth]{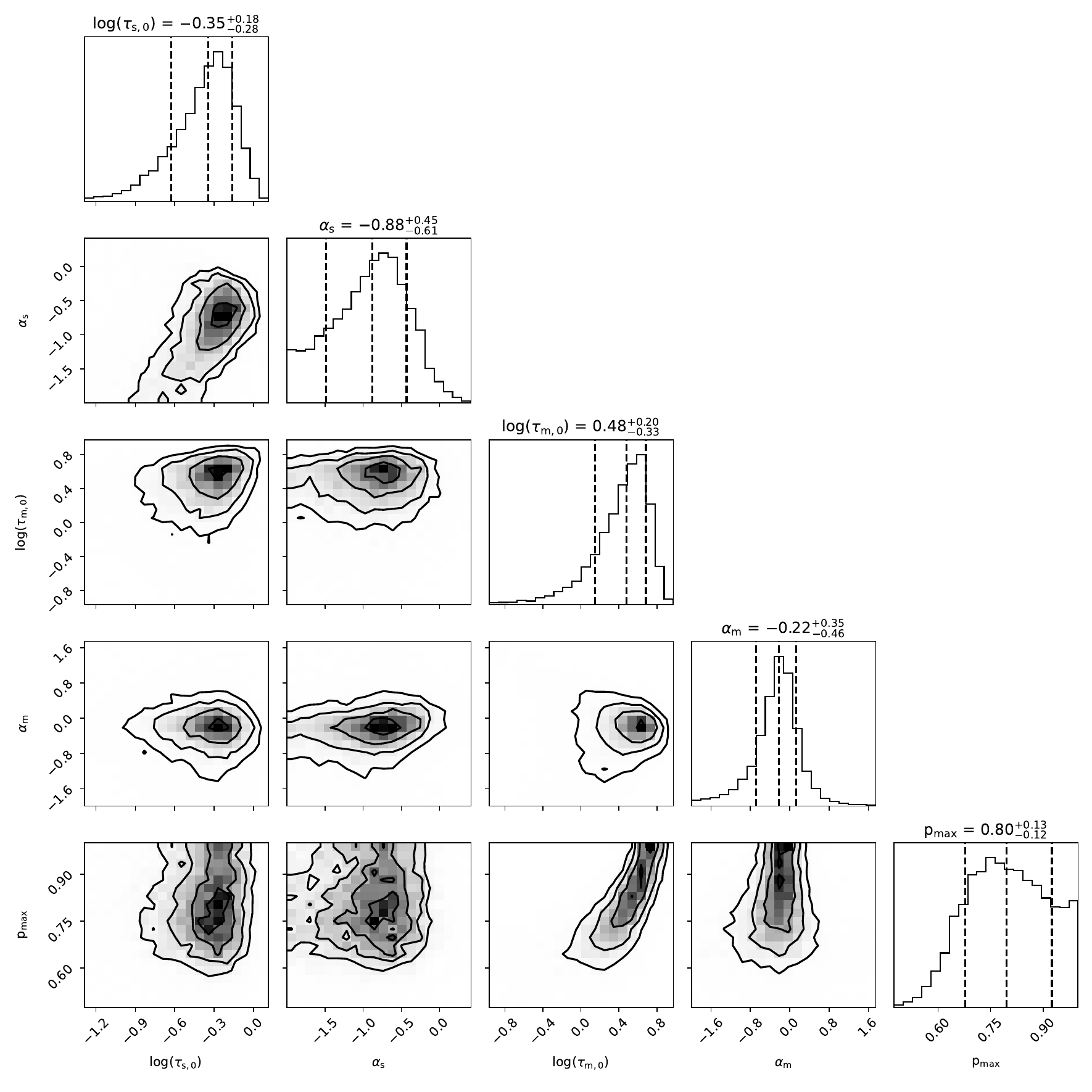}
\caption{Corner-plots showing the correlation between the various parameters (shaded regions, with contours at arbitrary iso-density levels) along with their marginalised probability distribution (histograms on top) for our model using the MS of \citet{Popesso+19}.}
\label{f:cornerplot_Popesso}
\end{center}
\end{figure*}

Fig.\,\ref{f:cornerplot_Popesso} shows the correlation between the various parameters for the model investigated in the main text.
The posteriors are all well behaved, but some degeneracy is present between $\tau_{\rm m}$ and $p_{\rm max}$.

Table \ref{t:param_Speagle} is equivalent to Table \ref{t:param_Popesso}, but is valid for models that use the MS of \citet{Speagle+14} instead of \citet{Popesso+23}.
The two MSs provide perfectly compatible fiducial parameters.

\begin{table}
\caption{Model parameters using the MS of \citet{Speagle+14}.}
\label{t:param_Speagle}
\centering
\begin{tabular}{lcccc}
\hline\hline
\noalign{\smallskip}
param. & ref. & prior & fiducial & best-fit \\
\hline
\noalign{\smallskip}
\multicolumn{5}{c}{cluster-driven quenching (RPS)}\\
\hline
\noalign{\smallskip}
$\log_{10}(\tau_{\rm s}/\rm Gyr)$ & Eq.\,(\ref{eq:ts}) & [-1.3,1.3] & $-0.33^{+0.18}_{-0.26}$ & $-0.24$\\
\noalign{\smallskip}
$\alpha_{\rm s}$ & Eq.\,(\ref{eq:ts}) & [-2.0,2.0] & $-0.95^{+0.45}_{-0.61}$ & $-0.83$\\
\hline
\noalign{\smallskip}
\multicolumn{5}{c}{cluster-driven morphological evolution}\\
\hline
\noalign{\smallskip}
$\log_{10}(\tau_{\rm m}/\rm Gyr)$ & Eq.\,(\ref{eq:tm}) & [-1.3,1.3] & $0.47^{+0.21}_{-0.33}$ & $0.66$\\
\noalign{\smallskip}
$\alpha_{\rm m}$ & Eq.\,(\ref{eq:tm}) & [-2.0,2.0] & $-0.23^{+0.37}_{-0.47}$ & $-0.20$\\
\noalign{\smallskip}
$p_{\rm max}$ & Eq.\,(\ref{eq:pS0}) & [0.0,1.0] & $0.79^{+0.13}_{-0.11}$ & $0.93$\\
\noalign{\smallskip}
\hline
\multicolumn{5}{p{0.45\textwidth}}{\small {\bf Column description:} (1)--parameter, (2)--reference Equation in the text, (3)--prior range, assuming uniform distributions, (4)--fiducial values from the 16th, 50th and 84th percentiles of the marginalised posterior distribution, (5)--values that maximise the likelihood.}
\end{tabular}
\end{table}

\end{appendix}
\end{document}